\documentclass[a4paper,11pt]{article} 

\usepackage[utf8]{inputenc}

\usepackage[english]{babel}
\usepackage{graphicx}          

\usepackage[toc,page]{appendix}
\usepackage{color}
\usepackage{dsfont}
\usepackage{extarrows}
\usepackage{enumerate}
\usepackage{amsmath}    
\usepackage{amssymb}
\usepackage{amsthm}
\usepackage{bbm}
\usepackage{graphicx}   
\usepackage{mathrsfs}
\usepackage{verbatim}   
\usepackage{subfigure}  
\usepackage{hyperref}   

\newtheorem{theorem}{Theorem}[section]
\newtheorem{corollary}[theorem]{Corollary}
\newtheorem{lemma}[theorem]{Lemma}
\newtheorem{proposition}[theorem]{Proposition}

\theoremstyle{definition}
\newtheorem{definition}[theorem]{Definition}
\newtheorem{remark}[theorem]{Remark}

\newtheorem*{problem*}{Problem}

\title{\LARGE \bf
Model robustness for feedback stabilization of open quantum systems
}
\date{\vspace{-5ex}}

\author{Weichao Liang,
\thanks{
{\small Department of Information Engineering, University of Padua, 6B, Via Gradenigo, 35131 Padova, Italy, (weichao.liang@dei.unipd.it).}}%
\and 
Nina H. Amini,
\thanks{
{\small Laboratoire des Signaux et Syst\`{e}mes (L2S), CNRS-CentraleSup\'{e}lec-Universit\'{e} Paris-Sud, Universit\'{e} Paris-Saclay, 3, Rue Joliot Curie, 91190 Gif-sur-Yvette, France, (nina.amini@l2s.centralesupelec.fr).}}
}

\begin{document}

\maketitle


\begin{abstract}                          
This paper generalizes the results in~\cite{liang2021robustness} concerning feedback stabilization of target states for $N$-level quantum angular momentum systems undergoing quantum non-demolition measurements (QND) in absence of the knowledge about initial states and parameters. Here we consider multiple measurement operators and study the stabilization toward a chosen target subspace which is a common eigenspace of measurement operators. Under the QND conditions, we show that this analysis provides necessary tools to ensure feedback stabilization based on a simplified filter whose state is a $N$-dimensional vector. A numerical analysis has been proposed in \cite{cardona2020exponential}. This paper provides a complete proof for the use of a simplified filter in feedback stabilization. This has important practical use as the dimension of quantum systems is usually high. This paper opens the way toward a complete proof concerning the robustness of a stabilizing feedback with respect to approximate filters, which is lacking. 
\end{abstract}

\section{Introduction}
The theory of open quantum systems~\cite{breuer2002theory}, that is systems which are in interaction with an external system (the environment or a bath), has direct impact in quantum information science~\cite{nielsen2010quantum}.
The interaction of open systems with environment causes quantum dissipation, i.e., the loss of information from the system to environment. The phenomenon is usually called decoherence. Compensation of such a decoherence is a principal issue in control theory of open quantum systems with technological applications~\cite{altafini2012modeling}. In particular, regarding robustness issues closed-loop control strategies are required.

The evolution of open quantum systems are described by quantum Langevin equations, obtained by quantum stochastic calculus \cite{hudson1984quantum}, quantum probability \cite{meyer2006quantum}, and input-output formalism~\cite{gardiner1985input}. The conditional evolution of an open quantum system undergoing indirect measurements is described by stochastic master equation. This is obtained by quantum filtering theory developed by Belavkin~\cite{belavkin1992quantum}.  
It is independently developed in physics community under the name quantum trajectory theory~\cite{carmichael2009open}. 
In principle indirect measurements, that is a system of interest $S$ indirectly measured through an auxiliary system which is in interaction with $S,$ is considered. Quantum measurements have probabilistic nature and have random back-action on the system, this has non-classical analogue. 
 
QND measurements (see~\cite{braginsky1995quantum} and also~\cite{haroche2006exploring} for further explanations) are of interest as they introduce less perturbations. For instance, the most standard measurements considered for feedback control though are QND. The first experiment stabilizing photon number states in the context of cavity quantum electrodynamics (QED) has been realized in \cite{sayrin2011real}. QND measurements allow non-deterministic preparation of pure states of measurement operators, this can be made deterministic by applying a feedback, see e.g., \cite{amini2012stabilization,amini2013feedback} for discrete-time and~\cite{van2005feedback,mirrahimi2007stabilizing,ticozzi2012stabilization,liang2019exponential} for continuous-time. 
In all of these papers, the control input appears in Lindblad generator through the system Hamiltonian. Theoretical parts concerning convergence and large time behaviors for quantum trajectories under QND measurements have been investigated in~\cite{bauer2013repeated,bauer2011convergence,benoist2014large}. In the context of circuit QED, the QND measurements and feedback have been considered firstly in~\cite{vijay2012stabilizing}.
 
In real experiments, there are many sources of imperfections, e.g., unknown initial states, detector inefficiency, unknown physical parameters, etc. Hence, robustness of control strategies is among predominant issues. In addition, a major difficulty with implementing a measurement-based feedback is that the slowest time scale is set by the classical processing of information (typically digitally) by quantum filters, which means that the feedback law is too slow for the system time-scale. In order to avoid fine-tuning of the noise parameters characterizing the decoherence processes, and avoid a dependence on initial conditions, model robustness has been studied in many papers, see e.g.,~\cite{liang2020robustness,liang2021robustness,lidar1998decoherence,ticozzi2008quantum}.
 
This paper demonstrates that our method in~\cite{liang2021robustness} can be adopted to solve more general problems.
We consider feedback stabilization of general open quantum systems undergoing QND measurements in presence of different measurement operators, not necessarily Hermitian (see e.g., \cite{sarlette2017deterministic}). Multiple measurement operators are particularly of interest for stabilization of entangled states, see e.g., \cite{zhang2020locally}. We obtain the convergence toward a target subspace, which is central in quantum error corrections where stabilization of a subspace of steady states is required, see e.g., \cite{ahn2002continuous,mabuchi2009continuous,cardona2019continuous}. Regarding stabilization of open quantum systems towards subspaces, in~\cite{benoist2017exponential}, the authors obtain results on exponential stabilization of subspaces with open-loop control strategies. This paper gives methods to feedback stabilization of subspaces for QND measurements in presence of imperfections, i.e., unknown initial states and physical parameters. Importantly, it is shown that the provided analysis can be applied to show the robustness of a stabilizing feedback depending only on an approximation of diagonal elements of the filter state with respect to the non-demolition basis. This means that the complexity of computing the evolution of filter state is reduced from $N^2$ to $N.$ This is important to pave the way for the use of approximate filter in feedback stabilization. To our knowledge, this presents the first  proof in this direction. In \cite{cardona2020exponential}, through a numerical analysis, the authors show the efficiency of a simplified filter in a feedback loop.  

 \smallskip
 
 This paper is structured as follows. In Section~\ref{Sec:SysDes}, we present the stochastic physical model, the main hypotheses on measurement operators, the notion of stochastic stability, and asymptotic properties of open-loop dynamics. Section \ref{sec:FullStateFeedback}
states a general theorem ensuring exponential stabilization of the target subspace for the coupled system~\eqref{Eq:SME_W}--\eqref{Eq:SME_filter_W} (Theorem \ref{Thm:Exp Stab General}). We give conditions ensuring instability of equilibria other than the target subspace and an estimation in mean of escaping time of the trajectories from a neighbourhood of such an invariant subspace (Proposition \ref{Prop:Instability}). This is later applied to provide sufficient conditions on the feedback controller and control Hamiltonian to ensure the recurrence property of the coupled system relative to the target subspace (Lemma~\ref{Lemma:Reachability}). This section concludes with application of Theorem \ref{Thm:Exp Stab General} giving sufficient conditions ensuring exponential stabilization (Theorem~\ref{Thm:ExpStab}). In Section \ref{sec:SimplifiedFeedback}, based on previous analysis, we prove that a simplified filter which depends only on an approximation of the diagonal populations of the estimated density matrix with respect to the non-demolition basis can still ensure the exponential stabilization of the system (Theorem \ref{Thm:ExpStab_q}). In Section \ref{sec:sim}, we provide simulations of a three-qubit system. This paper finishes in Section \ref{sec:conc} by giving a brief conclusion on this work.
\smallskip

\textbf{Notations.}
The imaginary unit is denoted by $i$. We take $\mathbf{I}$ as the identity operator on $\mathcal{H}$. We denote the adjoint $A\in\mathcal{B}(\mathcal{H})$ by $A^*$, where $\mathcal{B}(\mathcal{H})$ is the set of all linear operators on $\mathcal{H}$. The function $\mathrm{Tr}(A)$ corresponds to the trace of $A\in\mathcal{B}(\mathcal{H})$. The Hilbert-Schmidt norm of $A\in\mathcal{B}(\mathcal{H})$ is denoted by $\|A\|_{HS}:=\mathrm{Tr}(AA^*)^{1/2}$.
The commutator of two operators $A,B\in\mathcal{B}(\mathcal{H})$ is denoted by $[A,B]:=AB-BA.$ We denote by $\mathrm{int}(\mathcal{S})$ the interior of a subset of a topological space $\mathcal{S}$ and by $\partial \mathcal{S}$ its boundary. For $x\in\mathbb{C}$, $\mathbf{Re}\{x\}$ is the real part of $x$. 
For any finite positive integer $m$, denote $[m]:=\{1,\dots,m\}$.

\section{Stochastic dynamical model}
\label{Sec:SysDes}
We consider an open quantum system undergoing continuous-time measurements represented by a finite dimensional Hilbert space $\mathcal{H}$ with $\mathrm{dim}(\mathcal{H})=N$. We suppose we have $m$ different channels.
Assume that we do not have access to the initial state $\rho(0)$ and the actual physical parameters. In this case, we consider an estimation process with arbitrary initial state $\hat\rho(0)$ and physical parameters. The evolution of the actual quantum state and its associated estimated state can be described by the following stochastic master equations,
\begin{align}
d\rho(t)=&\mathcal{L}^u_{\omega,\gamma}(\rho(t))dt+\sum^m_{k=1}\mathcal{G}^k_{\eta_k, \gamma_k}(\rho(t))\Big(dY_k(t)\nonumber\\ 
&~~~~~~~~~~~~~~-\sqrt{\eta_k \gamma_k}\mathrm{Tr}\big((L_k+L^*_k)\rho(t)\big)dt\Big), \label{Eq:SME}
\\
d\hat{\rho}(t)=&\mathcal{L}^u_{\hat{\omega},\hat{\gamma}}(\hat\rho(t))dt+\sum^m_{k=1}\mathcal{G}^k_{\hat{\eta}_k, \hat{\gamma}_k}(\hat\rho(t))\Big(dY_k(t)\nonumber\\ 
&~~~~~~~~~~~~~~-\sqrt{\hat{\eta}_k \hat{\gamma}_k}\mathrm{Tr}\big((L_k+L^*_k)\hat{\rho}(t)\big)dt\Big),\label{Eq:SME_filter}
\end{align}
where
\begin{itemize}
\item the actual quantum state of the quantum system at time $t$ is denoted as $\rho(t)$, and belongs to the compact space $
\mathcal{S}:=\{\rho\in\mathcal{B}(\mathcal{H})|\,\rho=\rho^*\geq 0,\mathrm{Tr}(\rho)=1\}$. The associated estimated state is denoted as $\hat{\rho}(t)\in\mathcal{S}$;
\medskip
\item $\mathcal{L}^u_{\omega,\gamma}(\rho):=-i[\omega H_0+u H_1,\rho]+\sum^m_{k=1} F^k_{\gamma_k}(\rho)$ with $F^k_{\gamma_k}(\rho):=\gamma_k(L_k\rho L^*_k-L^*_kL_k \rho/2-\rho L^*_kL_k /2)$, and $\mathcal{G}^k_{\eta_k, \gamma_k}(\rho)=\sqrt{\eta_k}G^k_{\gamma_k}(\rho)$ with $G^k_{\gamma_k}(\rho):=\sqrt{\gamma_k}(L_k \rho+\rho L^*_k-\mathrm{Tr}((L_k+L^*_k)\rho)\rho)$ for all $k\in [m]$;
\medskip
\item $H_0=H^*_0\in\mathcal{B}(\mathcal{H})$ is the free Hamiltonian, $H_1=H^*_1\in\mathcal{B}(\mathcal{H})$ is the control Hamiltonian matrix corresponding to the action of external forces on the quantum system, $L_k\in\mathcal{B}(\mathcal{H})$ for all $k\in [m]$ are the measurement operators associated to $k$-th probe;
\medskip

\item $Y_k(t)$ denotes the observation process of $k$-th probe, which is a continuous semi-martingale whose quadratic variation is given by $\langle Y_k(t),Y_k(t)\rangle=t$. Denote the filtration generated by the observations up to time $t$ by $\mathcal{F}^{Y}_t:=\sigma(Y(s),0\leq s\leq t)$ where $Y(t)=(Y_k(t))_{1\leq k\leq m}$. Its dynamics satisfies $dY_k(t)=dW_k(t)+\sqrt{\eta_k \gamma_k}\mathrm{Tr}((L_k+L^*_k)\rho(t))dt$, where the innovation process $W_k(t)$ is a one-dimensional Wiener process and $\langle W_i(t),W_j(t)\rangle=\delta_{i,j}t$ for $i,j\in [m]$;
\medskip
\item $u:=u(\hat{\rho}) \in \mathbb{R}$ denotes the feedback controller adapted to $\mathcal{F}^{Y}_t$;
\medskip
\item $\omega\geq 0$ is a parameter characterizing the free Hamiltonian, $\gamma_k \geq 0$ is the strength of the interaction between the system and $k$-th probe\footnote{It should be noted that $\{\gamma_k\}$ can also be identified as the spectrum of the Gorini-Kossakowski-Sudarshan matrix~\cite{gorini1976completely,lindblad1976generators}.}, $\eta_k\in(0,1]$ describes the efficiency of the detector for $k$-th measurement channel. The estimated parameters $\hat{\omega} \geq 0$, $\hat{\gamma}_k \geq 0$ and $\hat{\eta}_k\in(0,1]$ may not be equal to the actual ones.
\end{itemize}
\medskip

By replacing $dY_k(t)=dW_k(t)+\sqrt{\eta_k \gamma_k}\mathrm{Tr}((L_k+L^*_k)\rho(t))dt$, we obtain the following matrix-valued stochastic differential equations in It\^o form in the filtered probability space $(\Omega,\mathcal{F},(\mathcal{F}_t),\mathbb{P})$, which are equivalent to~\eqref{Eq:SME}--\eqref{Eq:SME_filter}(see~\cite[Chapter 3]{barchielli2009quantum} for the detailed description). They describe the time evolution of the pair $(\rho(t),\hat{\rho}(t))\in\mathcal{S}\times\mathcal{S}$,
\begin{align}
d\rho(t)=\mathcal{L}^u_{\omega,\gamma}(\rho(t))dt+&\sum^m_{k=1}\mathcal{G}^k_{\eta_k, \gamma_k}(\rho(t))dW_k(t),\label{Eq:SME_W}\\
d\hat{\rho}(t)= \mathcal{L}^u_{\hat{\omega},\hat{\gamma}}(\hat\rho(t))dt+&\sum^m_{k=1}\mathcal{G}^k_{\hat{\eta}_k, \hat{\gamma}_k}(\hat\rho(t))\big(dW_k(t)+\mathcal{T}_k\big(\rho(t),\hat{\rho}(t)\big)dt\big),\label{Eq:SME_filter_W}
\end{align}
where $\mathcal{T}_k(\rho,\hat{\rho}):=\sqrt{\eta_k \gamma_k}\mathrm{Tr}((L_k+L^*_k)\rho)-\sqrt{\hat{\eta}_k \hat{\gamma}_k}\mathrm{Tr}((L_k+L^*_k)\hat{\rho})$.
The existence and uniqueness of the solution of~\eqref{Eq:SME_W}--\eqref{Eq:SME_filter_W} can be shown by similar arguments as in~\cite[Proposition 3.3]{mirrahimi2007stabilizing} and~\cite[Chapter 5]{barchielli2009quantum}. Moreover, it can be shown as in~\cite[Proposition 3.7]{mirrahimi2007stabilizing} that $(\rho(t),\hat{\rho}(t))$ is equipped with the Feller continuity and the strong Markov property in $\mathcal{S}\times\mathcal{S}$.

In this paper, we suppose that the measurement operators satisfy QND property. Hence there exists an orthogonal basis $\mathscr{B}:=\{|\phi_{k}\rangle\}^{N}_{k=1}$  of $\mathcal{H}$, where the system Hamiltonian $H_0$ and all measurement operators $L_k$ are diagonal in this basis (see e.g.,~\cite[Theorem 2]{benoist2014large}).  
It implies that we can write the measurement operators in the following form
$$
L_k =\sum^{N}_{n=1}\chi_{k,n} |\phi_{n}\rangle \langle \phi_{n}|  =\sum^{K_k}_{n=1}l_{k,n}\Pi_{k,n}, \quad  k\in [m],
$$
where $\chi_{k,n}\in\mathbb{C},$ and $l_{k,n}\in\mathbb{C}$ are the $K_k\leq N$ distinct eigenvalues of $L_k$ with the associated orthogonal projection $\Pi_{k,1},\dots,\Pi_{k,K_k}$ such that $\sum^{K_k}_{n=1}\Pi_{k,n}=\mathbf{I}$. 

\medskip

We impose the following assumption on the measurement operators which is needed later to ensure the exponential convergence when the feedback is turned off.
\smallskip
\begin{itemize}
\item[\textbf{H0}:] 
$\forall k\in [m]$, $\forall i,j\in [K_k]$ with $i\neq j$,  $\mathbf{Re}\{l_{k,i}\}\neq \mathbf{Re}\{l_{k,j}\}$.
\end{itemize}
\smallskip
For each $k\in [m]$, we set $\mathbf{Re}\{l_{k,1}\}<\dots<\mathbf{Re}\{l_{k,K_k}\}$. For the measurement operator $L_k$ with $k\in [m]$, we define $\underline{\ell}_{k}:=\min_{i\neq j}|\mathbf{Re}\{l_{k,i}\}-\mathbf{Re}\{l_{k,j}\}|>0$ for $i,j\in [K_k]$ and $\bar{\ell}_k:=|\mathbf{Re}\{l_{k,K_k}\}-\mathbf{Re}\{l_{k,1}\}|$.
Denote by $\{\mathcal{H}_n\}^M_{n=1}$ with $M\leq N$ the disjoint common eigenspaces of $H_0, L_1,\dots, L_m$ which are assumed to be non-empty. Let $P_n$ be the orthogonal projection on the eigenspace $\mathcal{H}_n$ for all $n\in [M]$. Hence, for any $n\in [M]$, $P_n$ is diagonal in the basis $\mathscr{B}$, $\sum^M_{n=1}P_n=\mathbf{I}$ and $P_iP_j=0$ for all $i\neq j$. For all $k\in [m]$ and $n\in [M]$, $L_k P_{n}=\mathfrak{l}_{k,n} P_{n}$ where $\mathfrak{l}_{k,n}\in\{l_{k,1},\dots,l_{k,K_k}\}$. 
Note that $\sum^m_{k=1}|\mathbf{Re}\{\mathfrak{l}_{k,i}\}-\mathbf{Re}\{\mathfrak{l}_{k,j}\}|\geq \min_{k\in[m]}\underline{\ell}_k$ with $i\neq j$, since the orthogonality of the projections $\{P_n\}^M_{n=1}$ and assumption \textbf{H0}.

To study the above problem, we first define different notions of stochastic stability for invariant subspaces of the coupled system. 
\subsection{Preliminary notions: invariant subspace and stochastic stability}
In this section, we define the notions of stochastic stability for an invariant subspace. To this end, we define $P_s\notin\{0,\mathbf{I}\}$ as the orthogonal projection on $\mathcal{H}_s\subset \mathcal{H}$ and the distance $\mathbf{d}_s(\rho):=\|\rho-P_s\rho P_s\|$ for $\rho\in\mathcal{S}$ and $P_s\rho P_s\in\mathcal B(\mathcal H_s),$ where $\|\cdot\|$ could be any matrix norm. 

\begin{lemma}
For all $\rho\in\mathcal{S}$ and any orthogonal projection $P$, $0\leq\mathrm{Tr}(\rho P)\leq 1$. Moreover, $\mathrm{Tr}(\rho P)=1$ is equivalent to $P\rho P=P\rho=\rho P=\rho.$
\label{Lemma:ProjTraceEquivalent}
\end{lemma}
The proof is based on simple linear algebra arguments and is omitted. Now denote the set of density matrices
\begin{equation*}
\mathcal{I}(\mathcal{H}_s):=\{\rho\in\mathcal{S}| \mathrm{Tr}(P_s\rho)=1\},
\end{equation*}
whose support is $\mathcal{H}_s$ or a subspace of $\mathcal{H}_s$.
\begin{definition}
The subspace $\mathcal{H}_s$ is called invariant almost surely if, for all $\rho(0)\in \mathcal{I}(\mathcal{H}_s)$, $\rho(t)\in \mathcal{I}(\mathcal{H}_s)$ for all $t>0$ almost surely.
\end{definition}
\textbf{Stochastic stability} 
Based on the stochastic stability defined in~\cite{khasminskii2011stochastic,mao2007stochastic} and the definition used in~\cite{ticozzi2008quantum,benoist2017exponential}, we formulate the following definition on the stochastic stability of the invariant subspace for the coupled system~\eqref{Eq:SME_W}--\eqref{Eq:SME_filter_W}.

\begin{definition}
Let $\mathcal{H}_s\times\mathcal{H}_s$ be the invariant subspace for~\eqref{Eq:SME_W}--\eqref{Eq:SME_filter_W} where $\mathcal{H}_s\subset\mathcal{H}$. Denote $P_s$ the orthogonal projection on $\mathcal{H}_s$. Then $\mathcal{H}_s\times\mathcal{H}_s$ is said to be
\begin{enumerate}
\item
\emph{Stable in probability}, if for every $\varepsilon \in (0,1)$ and for every $r >0$, there exists  $\delta = \delta(\varepsilon,r)>0$ such that,
\begin{equation*}
\begin{split}
\mathbb{P} \big( \mathbf{d}_s(\rho(t))+\mathbf{d}_s(\hat{\rho}(t))< r \text{ for } t \geq 0 \big) \geq 1-\varepsilon,
\end{split}
\end{equation*}
whenever $\mathbf{d}_s(\rho(0))+\mathbf{d}_s(\hat{\rho}(0))<\delta$ with $(\rho(0),\hat{\rho}(0))\in\mathcal{S}\times \mathcal{S}$. 

\item
\emph{Almost surely asymptotically stable}, if it is stable in probability and,
\begin{equation*}
\begin{split}
\mathbb{P} \big( \lim_{t\rightarrow\infty}\mathbf{d}_s(\rho(t))+\mathbf{d}_s(\hat{\rho}(t))=0 \big) = 1,
\end{split}
\end{equation*}
for all $(\rho(0),\hat{\rho}(0))\in\mathcal{S}\times \mathcal{S}$.

\item
\emph{Almost surely exponentially stable}, if
\begin{equation*}
\limsup_{t \rightarrow \infty} \frac{1}{t} \log \big( \mathbf{d}_s(\rho(t))+\mathbf{d}_s(\hat{\rho}(t))\big) < 0, \quad a.s.
\end{equation*}
for all $(\rho(0),\hat{\rho}(0))\in\mathcal{S}\times \mathcal{S}$. The left-hand side of the above inequality is called the \emph{sample Lyapunov exponent} of the solution.
\end{enumerate}
\end{definition}
In order to study the stabilization problem,
we introduce 
$
E_{s}(\rho):=\sqrt{1-\mathrm{Tr}(\rho P_s)}\in[0,1]
$
to estimate the above mentioned distance $\mathbf{d}_s(\rho)=\|\rho-P_s\rho P_s\|$ throughout of the paper.
Based on this function, we define
\begin{equation*}
\begin{split}
&B_{r}(\mathcal{H}_{s}):=\{\rho\in\mathcal{S}|\,E_{s}(\rho)<r\},\\
&\mathring{B}_{r}(\mathcal{H}_{s}):=\{\rho\in\mathcal{S}|\,0<E_{s}(\rho)<r\},\\
&\mathbf{B}_{r}(\mathcal{H}_{s}\times \mathcal{H}_{z}):=\{(\rho,\hat{\rho})\in\mathcal{S}\times \mathcal{S}|\,E_{s}(\rho)+E_{z}(\hat{\rho})<r\},\\
&\mathring{\mathbf{B}}_{r}(\mathcal{H}_{s}\times \mathcal{H}_{z}):=\{(\rho, \hat{\rho})\in\mathcal{S}\times \mathcal{S}|\,0<E_{s}(\rho)+E_{z}(\hat{\rho})<r\}.
\end{split}
\end{equation*} 

\begin{lemma}
For all $\rho\in\mathcal{S}$ and any orthogonal projection $P_s\in\mathcal{B}(\mathcal{H})$, there exist two constants $C_1>0$ and $C_2>0$ such that 
\begin{equation}
C_1E_{s}(\rho)^2\leq \|\rho-P_s\rho P_s\| \leq C_2 E_{s}(\rho).
\label{Eq:Relation_DisLya}
\end{equation}
\end{lemma}
\proof
By employing the arguments in the proof of~\cite[Lemma 4.8]{benoist2017exponential}, we have 
\begin{equation*}
\|\rho-P_s \rho P_s\|_{1}\leq N\|\rho-P_s \rho P_s\|_{\max}\leq 3N E_{s}(\rho)
\end{equation*}
where $\|\cdot\|_{1}$ and $\|\cdot\|_{\max}$ represent the trace norm and the max norm respectively. Moreover, since the trace norm is unitarily invariant, we have the following pinching inequality~\cite[Chapter 4.2]{bhatia2013matrix}
\begin{equation*}
\begin{split}
&\|\rho-P_s \rho P_s\|_{1}\\
&\geq\| P_s (\rho-P_s \rho P_s)P_s+(\mathbf{I}-P_s)(\rho-P_s \rho P_s)(\mathbf{I}- P_s)\|_{1}\\
&=\|(\mathbf{I}-P_s)\rho(\mathbf{I}-P_s)\|_{1}=E_{s}(\rho)^2.
\end{split}
\end{equation*}
Then, the equivalence of matrix norms on finite dimensional vector spaces concludes the proof.
\hfill$\square$

\subsection{Asymptotic properties of open-loop dynamics}
The asymptotic behavior of the solution of~\eqref{Eq:SME_W} in open-loop case has been studied in several papers (see e.g.,~\cite{benoist2014large,liang2019exponential,liang2022GHZ,cardona2020exponential}), which is known as exponential quantum state reduction.
It plays a fundamental role in stabilization strategies. Here we show an exponential quantum state reduction for the actual system with providing explicit convergence rate. We make use of the convergence rate information imposed by the diffusion term of the stochastic master equations~\eqref{Eq:SME_W} to render the trajectories of the coupled system~\eqref{Eq:SME_W}--\eqref{Eq:SME_filter_W} toward the target subspace.
In this case, for all $n\in [M],$  it can be easily verified that $\mathcal{H}_n$ is invariant almost surely for the system~\eqref{Eq:SME_W} (see~\cite[Theorem 1.1]{benoist2017exponential} for the proof). For the reader convenience, we recall the following result.
\begin{theorem}[Exponential quantum state reduction]
Assume that \textbf{\emph{H0}} holds. For System~\eqref{Eq:SME_W} with $\rho(0) \in \mathcal{S}$,  $\{\mathcal{H}_n\}^{M}_{n=1}$ is exponentially stable in mean and a.s. with average and sample Lyapunov exponent less than or equal to $-\frac{1}{2}\min_{k\in [m]}\{\eta_k \gamma_k \underline{\ell}_k^2\}$. Moreover, the probability of convergence to $\mathcal H_n$ is $\mathrm{Tr}(\rho(0) P_n).$
\label{Thm:QSR}
\end{theorem}
This can be proved by using similar arguments as in~\cite[Theorem 5]{liang2019exponential} or~\cite[Theorem 3]{liang2022GHZ} and by consideing the candidate Lyapunov function $V(\rho)=\sum^M_{i\neq j}\sqrt{\mathrm{Tr}(\rho P_i)\mathrm{Tr}(\rho P_j)}$. We can show that  
\begin{equation*}
\mathscr{L} V(\rho)\leq -\frac{1}{2}\min_{k\in [m]}\{\eta_k \gamma_k \underline{\ell}_k^2\} V(\rho).
\end{equation*}
It implies that the $V(\rho(t))$ in open-loop case converges exponentially to zero in mean and almost surely with the Lyapunov exponent less than or equal to $-\frac{1}{2}\min_{k\in [m]}\{\eta_k \gamma_k \underline{\ell}_k^2\}$. It means that $\rho(t)$ in open-loop case converges exponentially to one of $\{\mathcal{I}(\mathcal{H}_n)\}^M_{n=1}$ with certain probability. The probability of convergence to $\mathcal H_n$ can be precised by similar arguments provided in ~\cite[Theorem 5]{liang2019exponential} or~\cite[Theorem 3]{liang2022GHZ}.

\begin{remark}
Rather than constructing a Lyapunov function to analyze the large time behaviour of system~\eqref{Eq:SME_W}, in~\cite{benoist2014large}, the authors study the dynamics of $\mathrm{Tr}(\rho(t)P_n)$ with $n\in [M]$ in Dol\'eans-Dade exponential form, which leads to a more precise estimation of Lyapunov exponent. Furthermore, the authors show that in the open-loop case when $(\omega,\{\gamma_k\},\{\eta_k\})=(\hat{\omega},\{\hat{\gamma}_k\},\{\hat{\eta}_k\})$ and $\rho(0)\neq \hat{\rho}(0)$, under the assumption that $\mathrm{Tr}(\hat{\rho}(0)P_n)>0$ for any $n\in [M]$ such that $\mathrm{Tr}(\rho(0)P_n)> 0$, $\rho(t)$ and $\hat{\rho}(t)$ converge exponentially to the same subset $\mathcal{I}(\mathcal{H}_{n})$ with the same probability. Note that if all the common eigenspace $\{\mathcal{H}_n\}^M_{n=1}$ are one-dimensional, then $M=N$, and in this case $\rho(t)$ and $\hat{\rho}(t)$ converge exponentially to $|\Upsilon\rangle\langle\Upsilon|=\sum^N_{n=1} P_n \mathrm{Tr}(\rho(0)P_n)$, the case established in \cite{benoist2014large}. However, this method cannot treat the case where $(\omega,\{\gamma_k\},\{\eta_k\})\neq(\hat{\omega},\{\hat{\gamma}_k\},\{\hat{\eta}_k\})$. In open-loop case, the exponential stability for the case of unknown parameters will be appeared in~\cite{liang2022robust}.
\label{Rem:QSR}
\end{remark}

\section{Feedback stabilization toward target subspaces}
\label{sec:FullStateFeedback}
In this section, we provide an adaptation of~\cite[Theorem 4.11]{liang2021robustness} ensuring exponential stabilization of the target subspace $\mathcal{H}_{\bar{n}}\times \mathcal{H}_{\bar{n}}$. It should be noted that this is inspired by~\cite[Theorem 2.12]{baxendale1991invariant},~\cite[Theorem 5.7]{khasminskii2011stochastic} and~\cite[Theorem 3.3, 2.3]{mao2007stochastic}.   
\begin{theorem}
Suppose that the trajectories $(\rho(t), \hat{\rho}(t))$ are recurrent relative to any neighborhood of $\mathcal{I}(\mathcal{H}_{\bar{n}})\times\mathcal{I}(\mathcal{H}_{\bar{n}})$. Additionally, assume the existence of a positive-definite function $V(\rho,\hat{\rho})$ such that $V(\rho,\hat{\rho})\!\!=\!\!0$ if and only if $(\rho,\hat{\rho})\in\mathcal{I}(\mathcal{H}_{\bar{n}})\times\mathcal{I}(\mathcal{H}_{\bar{n}})$, and $V$ is continuous on $\mathcal{S}\times\mathcal{S}$ and twice continuously differentiable on an almost surely invariant subset $\Gamma$ of $\mathcal{S}\times\mathcal{S}$ containing $\mathrm{int}(\mathcal{S})\times\mathrm{int}(\mathcal{S})$.
Moreover, suppose that there exist positive constants $C$, $C_1$ and $C_2$ such that 
\begin{enumerate}
\item[(i)] $C_1 \, \big(E_{\bar{n}}(\rho)+E_{\bar{n}}(\hat{\rho})\big)^{C_2}\leq V(\rho,\hat{\rho})$, for all $(\rho,\hat{\rho})\in\mathcal{S}\times\mathcal{S}$, and 
\item[(ii)] $\limsup_{(\rho,\hat{\rho})\rightarrow\mathcal{I}(\mathcal{H}_{\bar{n}})\times\mathcal{I}(\mathcal{H}_{\bar{n}})}\frac{\mathscr{L}V(\rho,\hat{\rho})}{V(\rho,\hat{\rho})}\leq-C$.
\end{enumerate}
Then, $\mathcal{H}_{\bar{n}}\times\mathcal{H}_{\bar{n}}$ is almost surely exponentially stable for~\eqref{Eq:SME_W}--\eqref{Eq:SME_filter_W} starting from $\Gamma$ with sample Lyapunov exponent less than or equal to $-(C+K)/C_2$, where $K\leq \liminf_{(\rho,\hat{\rho})\rightarrow\mathcal{I}(\mathcal{H}_{\bar{n}})\times\mathcal{I}(\mathcal{H}_{\bar{n}})}\frac{1}{2}\sum^m_{k=1}g_k(\rho,\hat{\rho})^2$ and 
$$g_k(\rho,\hat{\rho})\!:=\!\mathrm{Tr}\Big(\frac{\partial V(\rho,\hat{\rho})}{\partial \rho}\frac{\mathcal{G}^k_{\eta_k,\gamma_k}(\rho)}{ V(\rho,\hat{\rho})}+\frac{\partial V(\rho,\hat{\rho})}{\partial \hat{\rho}}\frac{\mathcal{G}^k_{\hat{\eta}_k,\hat{\gamma}_k}(\hat{\rho})}{ V(\rho,\hat{\rho})}\Big).\footnote{Here $\frac{\partial V}{\partial\rho}$ denotes the first differential in $\rho\in\mathcal{S}$ of the function $V$, which can be expanded in terms of partial derivatives as in~\eqref{Eq:InfinitesimalGenerator}.}$$
\label{Thm:Exp Stab General}
\end{theorem}
The proof can be done by showing first the asymptotic stability and then estimating the sample Lyapunov exponent. The ideas are resumed as follows.
\medskip

\begin{enumerate}
\item (Asymptotic stability) Together with the recurrence property and the strong Markov property of $(\rho(t),\hat{\rho}(t))$, the stability in probability implies the almost sure asymptotic stability of $\mathcal{I}(\mathcal{H}_{\bar{n}})\times \mathcal{I}(\mathcal{H}_{\bar{n}})$.
\item (Exponential stability) We provide an estimation of the sample Lyapunov exponent. 
\end{enumerate}
By slightly modifying~\cite[Theorem 5.7]{khasminskii2011stochastic}, we obtain the following result. Denote $\mathcal{K}$ as the family of all continuous non-decreasing functions $\mu:\mathbb{R}_{\geq 0}\rightarrow\mathbb{R}_{\geq0}$ such that $\mu(0)=0$ and $\mu(r)>0$ for all $r>0$.
\begin{proposition}
Suppose that $(\rho(t),\hat{\rho}(t))$ is recurrent relative to any neighborhood of $\mathcal{I}(\mathcal{H}_{\bar{n}})\times \mathcal{I}(\mathcal{H}_{\bar{n}})$. 
Additionally, suppose that there exists a positive-definite function $V(\rho,\hat{\rho})$ such that $V(\rho,\hat{\rho})=0$ if and only if $(\rho,\hat{\rho})=\mathcal{I}(\mathcal{H}_{\bar{n}})\times\mathcal{I}(\mathcal{H}_{\bar{n}})$, and $V$ is continuous on $\mathcal{S}\times\mathcal{S}$ and twice continuously differentiable on an almost surely invariant subset $\Gamma$ of $\mathcal{S}\times\mathcal{S}$ containing $\mathrm{int}(\mathcal{S})\times\mathrm{int}(\mathcal{S})$.
Moreover, suppose that there exists a function $\mu\in\mathcal{K}$ such that 
\begin{enumerate}
\item[(i)] $V(\rho,\hat{\rho})\geq \mu\big(E_{\bar{n}}(\rho)+E_{\bar{n}}(\hat{\rho})\big)$, for all $(\rho,\hat{\rho})\in\mathcal{S}\times\mathcal{S}$, and 
\item[(ii)] $\mathscr{L}V(\rho,\hat{\rho})\leq0$ for all $(\rho,\hat{\rho})\in\mathbf{B}_r(\mathcal{H}_{\bar{n}}\times\mathcal{H}_{\bar{n}})$ with some $r>0$.
\end{enumerate}
Then, $\mathcal{H}_{\bar{n}}\times\mathcal{H}_{\bar{n}}$ is almost surely asymptotically stable for the coupled system~\eqref{Eq:SME_W}--\eqref{Eq:SME_filter_W} starting from $\Gamma.$
\label{Cor:Asym Stab General}
\end{proposition}

In the following, we provide sufficient conditions ensuring that $(\rho(t),\hat{\rho}(t))$ is recurrent relative to any neighborhood of $\mathcal{I}(\mathcal{H}_{\bar{n}})\times \mathcal{I}(\mathcal{H}_{\bar{n}})$, i.e., $(\rho(t),\hat{\rho}(t))$ enters any neighborhood of $\mathcal{I}(\mathcal{H}_{\bar{n}})\times \mathcal{I}(\mathcal{H}_{\bar{n}})$ in finite time almost surely (\hyperref[Lemma:Reachability]{Lemma~\ref*{Lemma:Reachability}}). To this end, we first provide sufficient conditions ensuring the exit of $(\rho(t),\hat{\rho}(t))$ from a neighborhood of any invariant subset $\mathcal{I}(\mathcal{H}_{n})\times \mathcal{I}(\mathcal{H}_{\bar{n}})$ with $n\neq \bar{n}$ in finite time almost surely. Finally, we provide examples giving sufficient conditions ensuring exponential
stabilization (\hyperref[Thm:ExpStab]{Theorem~\ref*{Thm:ExpStab}}). 
\begin{remark}We note that these main steps are fundamental as in the the proof of~\cite[Theorem 4.11]{liang2021robustness}, however the technical details are different as we work with different measurement channels and we stabilize target subspacs, with particular interests in entangled states stabilization and quantum error corrections. More originally, we apply theses analyses to prove that a partial information on the state is sufficient  to prove exponential stabilization which is lacking in the literature (see Section \ref{sec:SimplifiedFeedback}).
\end{remark}

\subsection{Sufficient conditions ensuring recurrence}\label{sec:general}
Here we provide sufficient conditions on the feedback controller and control Hamiltonian to ensure the recurrence of $(\rho(t),\hat{\rho}(t))$ relative to any neighborhood of $\mathcal{I}(\mathcal{H}_{\bar{n}})\times \mathcal{I}(\mathcal{H}_{\bar{n}})$, i.e., for all $(\rho(0),\hat{\rho}(0))\in\mathcal{S}\times \mathcal{S} \setminus \{\bigcup_{n\neq\bar{n}}\mathcal{I}(\mathcal{H}_n)\times\mathcal{I}(\mathcal{H}_{\bar{n}})\}$, $(\rho(t),\hat{\rho}(t))$ can arbitrarily approach $\mathcal{I}(\mathcal{H}_{\bar{n}})\times\mathcal{I}(\mathcal{H}_{\bar{n}})$ in finite time almost surely. To prove this, we provide conditions to show instability of other invariant subspaces and the reachability for the deterministic control system. 
\subsubsection{Instability of $\mathcal{I}(\mathcal{H}_n)\times\mathcal{I}(\mathcal{H}_{\bar{n}})$ with $n\neq\bar{n}$} 

Inspired by Khas'minskii's recurrence conditions~\cite[Theorem 3.9]{khasminskii2011stochastic}, in the following proposition, we provide Lyapunov-type sufficient conditions ensuring the instability of an invariant subspace of the coupled system~\eqref{Eq:SME_W}--\eqref{Eq:SME_filter_W} and an estimation in mean of escaping time of the trajectories from a neighbourhood of such invariant subspace. Since the proof is based on the similar arguments as in~\cite[Theorem 3.9]{khasminskii2011stochastic}, we omit it. We consider the stopping time $\tau^{n,k}_{\lambda}:=\inf\{t\geq0|\,E_n(\rho(t))+E_k(\hat{\rho}(t))=\lambda\}$ with $n,k\in[M],$ $\lambda>0$, and setting $\inf\{\emptyset\}=\infty$.
\begin{proposition}
Consider an invariant subspace $\mathcal{H}_n\times\mathcal{H}_k$ of the coupled system~\eqref{Eq:SME_W}--\eqref{Eq:SME_filter_W} with $n,k\in[M]$. Suppose that there exist $V\in\mathcal{C}^2(\mathcal{S}\times\mathcal{S},\mathbb{R}_{\geq 0})$, $C>0$ and $\lambda\in(0,2)$ such that $\mathscr{L}V(\rho,\hat{\rho})\leq -C$ whenever $(\rho,\hat{\rho})\in\mathbf{B}_{\lambda}(\mathcal{H}_{n}\times\mathcal{H}_{k})$. Then, for any $(\rho(0),\hat{\rho}(0))\in\mathring{\mathbf{B}}_{\lambda}(\mathcal{H}_{n}\times\mathcal{H}_{k})$, we have $\mathbb{E}(\tau^{n,k}_{\lambda})<\frac{1}{C}V(\rho(0),\hat{\rho}(0))$.
\label{Prop:Instability}
\end{proposition}
Based on \hyperref[Prop:Instability]{Proposition~\ref*{Prop:Instability}} and techniques of estimating the lower bound of Lyapunov exponents in~\cite[Theorem 3.5]{mao2007stochastic}, in the following lemmas, we show the instability of the subsets $\mathcal{I}(\mathcal{H}_n)\times\mathcal{I}(\mathcal{H}_{\bar{n}})$ with $n\neq\bar{n}$ for two cases separately, and provide an estimation of the divergence rate of $(\rho(t),\hat{\rho}(t))$ from $\mathcal{I}(\mathcal{H}_n)\times\mathcal{I}(\mathcal{H}_{\bar{n}})$.

For all $k\in[M]$ and $n\in[M]$, let $L_k P_{n}=\mathfrak{l}_{k,n} P_{n}$ where $\mathfrak{l}_{k,n}\in\{l_{k,1},\dots,l_{k,K_k}\}$. Define 
$
\overline{C}_{k,\bar{n}}:=\mathbf{Re}\{\mathfrak{l}_{k,\bar{n}}\}-\min_{n\neq \bar{n}}\mathbf{Re}\{\mathfrak{l}_{k,n}\}$, $\underline{C}_{k,\bar{n}}:=\mathbf{Re}\{\mathfrak{l}_{k,\bar{n}}\}-\max_{n\neq \bar{n}}\mathbf{Re}\{\mathfrak{l}_{k,n}\}$.
Furthermore, for all $n\in[M]$ and $k\in[M]$, we set $\Theta_n(\rho):=\mathrm{Tr}(i[H_1,{\rho}]P_n)$, $\Delta_{k,n}(\rho):=2\mathbf{Re}\{\mathfrak{l}_{k,n}\}-\mathrm{Tr}((L_k+L_k^*)\rho)$, $\hat{\mathsf{G}}^{n}_{k}(\hat{\rho}):=\mathrm{Tr}\big(\mathcal{G}^{k}_{\hat{\eta}_k,\hat{\gamma}_k}(\hat{\rho})P_{n}\big)$, and define 
\begin{equation*}
\begin{split}
\overline{T}_{k,\bar{n}}&:=\sqrt{\eta_k \gamma_k}\big(\mathbf{Re}\{\mathfrak{l}_{k,\bar{n}}\}-\underline{C}_{k,\bar{n}}\big)-\sqrt{\hat{\eta}_k \hat{\gamma}_k}\mathbf{Re}\{\mathfrak{l}_{k,\bar{n}}\},\\
\underline{T}_{k,\bar{n}}&:=\sqrt{\eta_k \gamma_k}\big(\mathbf{Re}\{\mathfrak{l}_{k,\bar{n}}\}-\overline{C}_{k,\bar{n}}\big)-\sqrt{\hat{\eta}_k \hat{\gamma}_k}\mathbf{Re}\{\mathfrak{l}_{k,\bar{n}}\}.
\end{split}
\end{equation*}
Based on above definitions, we define partitions of $[M]$,
$\overline{\mathcal{K}}^+_{\bar{n}}:=\{k\in[M]|\,\underline{T}_{k,\bar{n}}\geq0\}$, $\mathcal{K}^+_{\bar{n}}:=\{k\in \overline{\mathcal{K}}^+_{\bar{n}}|\, \overline{C}_{k,\bar{n}}<0\}$, $\overline{\mathcal{K}}^-_{\bar{n}}:=\{k\in[M]|\,\overline{T}_{k,\bar{n}}\leq0\}$, $\mathcal{K}^-_{\bar{n}}:=\{k\in \overline{\mathcal{K}}^-_{\bar{n}}|\, \underline{C}_{k,\bar{n}}>0\}$.
Next, we introduce the following assumptions on feedback controller:
\smallskip
\begin{itemize}
\item[\textbf{H1}:] 
$|u(\hat{\rho})|\leq c(1-\mathrm{Tr}(\hat{\rho}P_{\bar{n}}))^\alpha$ with $\alpha>1/2$ and $c>0$.
\item[\textbf{H2}:] 
$u(\hat{\rho})=0$ for all $\hat{\rho}\in B_{\epsilon}(\mathcal{H}_{\bar{n}})$ with $\epsilon>0$.
\end{itemize}
\smallskip
The above assumptions are needed to show the instability of $\mathcal{I}(\mathcal{H}_n)\times\mathcal{I}(\mathcal{H}_{\bar{n}})$ with $n\neq\bar{n}$ and the reachability of $\mathcal{I}(\mathcal{H}_{\bar{n}})\times\mathcal{I}(\mathcal{H}_{\bar{n}})$. 
\begin{lemma}
Assume that \emph{\textbf{H0}} and  \emph{\textbf{H1}} hold. Suppose that $\overline{\mathcal{K}}^+_{\bar{n}}\cup \overline{\mathcal{K}}^-_{\bar{n}}=[M]$ with $\bar{n}\in[M]$ and 
\begin{equation}
D_{\bar{n}}>\sum^{m}_{k=1}\hat{\eta}_k \hat{\gamma}_k \max\{\overline{C}_{k,\bar{n}}^2,\underline{C}_{k,\bar{n}}^2\}
\label{Eq:ConditionInstability_S1}
\end{equation}
with  
\begin{equation*}
\begin{split}
&D_{\bar{n}}:=\sum_{k\in \mathcal{K}^+_{\bar{n}}}\sqrt{\hat{\eta}_k \hat{\gamma}_k} |\overline{C}_{k,\bar{n}}|\underline{T}_{k,\bar{n}}+\sum_{k\in \mathcal{K}^-_{\bar{n}}}\sqrt{\hat{\eta}_k \hat{\gamma}_k} \underline{C}_{k,\bar{n}}|\overline{T}_{k,\bar{n}}|\\
&-\sum_{k\in \overline{\mathcal{K}}^+_{\bar{n}} \setminus \mathcal{K}^+_{\bar{n}}}\sqrt{\hat{\eta}_k \hat{\gamma}_k} \overline{C}_{k,\bar{n}}\overline{T}_{k,\bar{n}}-\sum_{k\in \overline{\mathcal{K}}^-_{\bar{n}} \setminus \mathcal{K}^-_{\bar{n}}}\sqrt{\hat{\eta}_k \hat{\gamma}_k} \underline{C}_{k,\bar{n}}\underline{T}_{k,\bar{n}}.
\end{split}
\end{equation*}
Then, there exists $\lambda>0$ such that, for all initial condition $(\rho(0),\hat{\rho}(0))\in\mathbf{B}_{\lambda}(\mathcal{H}_{n}\times\mathcal{H}_{\bar{n}})$ with $n\neq\bar{n}$, the solutions $(\rho(t),\hat{\rho}(t))$ of the coupled system~\eqref{Eq:SME_W}--\eqref{Eq:SME_filter_W} exit $\mathbf{B}_{\lambda}(\mathcal{H}_{n}\times\mathcal{H}_{\bar{n}})$ in finite time almost surely. 
\label{Lemma:Unstability_Special}
\end{lemma}

\proof
To prove the lemma, we apply \hyperref[Prop:Instability]{Proposition~\ref*{Prop:Instability}} with the Lyapunov function $-\log \big(1-\mathrm{Tr}(\hat{\rho}P_{\bar{n}})\big)\geq 0$. Set $V_{\bar{n}}(\hat{\rho}):=1-\mathrm{Tr}(\hat{\rho}P_{\bar{n}})\in[0,1]$. The assumptions of \hyperref[Lemma:NeverReach]{Lemma~\ref*{Lemma:NeverReach}} hold due to \textbf{H1}, the infinitesimal generator of the Lyapunov function is given by
\begin{equation*}
\mathscr{L}-\log V_{\bar{n}}(\hat{\rho})=-\frac{\mathscr{L} V_{\bar{n}}(\hat{\rho})}{V_{\bar{n}}(\hat{\rho})}+\frac{1}{2}\frac{1}{V_{\bar{n}}(\hat{\rho})^2}\sum^{m}_{k=1}\big(\hat{\mathsf{G}}^{\bar{n}}_{k}(\hat{\rho})\big)^2,
\end{equation*}
where $\hat{\mathsf{G}}^{\bar{n}}_{k}(\hat{\rho})=\sqrt{\hat{\eta}_k\hat{\gamma}_k}\Delta_{k,\bar{n}}(\hat{\rho})\mathrm{Tr}(\hat{\rho}P_{\bar{n}})$ and
\begin{equation*}
\mathscr{L} V_{\bar{n}}(\hat{\rho})=u\Theta_{\bar{n}}(\hat{\rho})-\sum^m_{k=1}\sqrt{\hat{\eta}_k\hat{\gamma}_k}\Delta_{k,\bar{n}}(\hat{\rho})\mathcal{T}_k(\rho,\hat{\rho})\mathrm{Tr}(\hat{\rho}P_{\bar{n}}).
\end{equation*}
By \textbf{H1}, we have $|u|\leq c \big(1-\mathrm{Tr}(\hat{\rho}P_{\bar{n}})\big)^\alpha$ with $\alpha>1/2$ and $c>0$. Moreover,
\begin{equation*}
\begin{split}
&|\mathrm{Tr}(i[H_1,\hat{\rho}]P_{\bar{n}})|\\
&=|\mathrm{Tr}(i[H_1,\hat{\rho}](P_{\bar{n}}-\mathbf{I}))|\leq |\mathrm{Tr}(H_1\hat{\rho}(P_{\bar{n}}-\mathbf{I}))|+|\mathrm{Tr}(H_1(P_{\bar{n}}-\mathbf{I})\hat{\rho})|\\
&\leq 2\|H_1\sqrt{\hat{\rho}}\|_{HS}\|(P_{\bar{n}}-\mathbf{I})\sqrt{\hat{\rho}}\|_{HS}=2\sqrt{2}\|H_1\sqrt{\hat{\rho}}\|_{HS}\sqrt{1-\mathrm{Tr}(\hat{\rho}P_{\bar{n}})}.
\end{split}
\end{equation*}
Since $\{P_n\}^M_{n=1}$ resolve the identity, for all $k\in[M]$, we have the following inequalities
\begin{equation}
\underline{C}_{k,\bar{n}}\big(1-\mathrm{Tr}(\hat{\rho}P_{\bar{n}})\big)\leq \frac{\Delta_{k,\bar{n}}(\hat{\rho})}{2}\leq \overline{C}_{k,\bar{n}}\big(1-\mathrm{Tr}(\hat{\rho}P_{\bar{n}})\big).
\label{Eq:IneqLambda}
\end{equation}
It is easy to verify that, for all $k\in[M]$, $\mathcal{T}_{k}(\rho,\hat{\rho})>0$ if $\underline{T}_{k,\bar{n}}>0$ and $\mathcal{T}_{k}(\rho,\hat{\rho})<0$ if $\overline{T}_{k,\bar{n}}<0$, when $(\rho,\hat{\rho})$ is close enough to $\mathcal{I}(\mathcal{H}_{n})\times\mathcal{I}(\mathcal{H}_{\bar{n}})$ with $n\neq \bar{n}$. 
Hence, in such a neighbourhood, 
\begin{equation*}
\begin{split}
\mathscr{L} V_{\bar{n}}(\hat{\rho}) \geq \Big(&-c V_{\bar{n}}(\hat{\rho})^{\alpha-\frac{1}{2}}+2\sum_{k\in \overline{\mathcal{K}}^-_{\bar{n}}}\sqrt{\hat{\eta}_k\hat{\gamma}_k}\overline{C}_{k,\bar{n}}\mathcal{T}_k(\rho,\hat{\rho})\\
&+2\sum_{k\in \overline{\mathcal{K}}^+_{\bar{n}}}\sqrt{\hat{\eta}_k\hat{\gamma}_k}\underline{C}_{k,\bar{n}}\mathcal{T}_k(\rho,\hat{\rho})\Big) V_{\bar{n}}(\hat{\rho})
\end{split}
\end{equation*} 
It implies that 
$
\liminf_{(\rho,\hat{\rho})\rightarrow \mathcal{I}(\mathcal{H}_{n})\times \mathcal{I}(\mathcal{H}_{\bar{n}})}\frac{\mathscr{L}V_{\bar{n}}(\hat{\rho})}{V_{\bar{n}}(\hat{\rho})}\geq 2D_{\bar{n}}.
$
Moreover, due to the inequality 
\begin{equation*}
\frac{\sum^{m}_{k=1}\big(\hat{\mathsf{G}}^{\bar{n}}_{k}(\hat{\rho})\big)^2}{V_{\bar{n}}(\hat{\rho})^2} \leq 4\sum^{m}_{k=1} \hat{\eta}_k \hat{\gamma}_k \max\{\overline{C}_{k,\bar{n}}^2,\underline{C}_{k,\bar{n}}^2\}\mathrm{Tr}(\hat{\rho}P_{\bar{n}})^2,
\end{equation*}
we can deduce that
\begin{equation*}
\begin{split}
\limsup_{(\rho,\hat{\rho})\rightarrow \mathcal{I}(\mathcal{H}_{n})\times \mathcal{I}(\mathcal{H}_{\bar{n}})}&\frac{1}{2}\frac{1}{V_{\bar{n}}(\hat{\rho})^2}\sum^{m}_{k=1}\big(\hat{\mathsf{G}}^{\bar{n}}_{k}(\hat{\rho})\big)^2\leq 2\sum^{m}_{k=1} \hat{\eta}_k \hat{\gamma}_k \max\{\overline{C}_{k,\bar{n}}^2,\underline{C}_{k,\bar{n}}^2\}.
\end{split}
\end{equation*}
The condition~\eqref{Eq:ConditionInstability_S1} guarantees that there exist $C>0$ and $\lambda\in(0,2)$ such that, for all $n\neq \bar{n}$, 
\begin{equation*}
\mathscr{L}-\log V_{\bar{n}}(\hat{\rho})\leq -C, \quad \forall (\rho,\hat{\rho})\in\mathbf{B}_{\lambda}(\mathcal{H}_{n}\times\mathcal{H}_{\bar{n}}).
\end{equation*}
Therefore, by applying \hyperref[Prop:Instability]{Proposition~\ref*{Prop:Instability}}, for any $(\rho(0),\hat{\rho}(0))\in\mathring{\mathbf{B}}_{\lambda}(\mathcal{H}_{n}\times\mathcal{H}_{k})$, we have $\mathbb{E}(\tau^{n,\bar{n}}_{\lambda})<-\frac{1}{C}\log V_{\bar{n}}(\hat{\rho}(0))<\infty$. Then, by Markov inequality, we have $\mathbb{P}(\tau^{n,\bar{n}}_{\lambda}<\infty)=1$.
\hfill$\square$

\smallskip

Now, we show the instability of $\mathcal{I}(\mathcal{H}_{n})\times \mathcal{I}(\mathcal{H}_{\bar{n}})$ with $n\neq \bar{n}$ by the following lemma, where the assumption of \hyperref[Lemma:Unstability_Special]{Lemma~\ref*{Lemma:Unstability_Special}} are not necessarily verified. 
\begin{lemma}
Let $\bar{n}\in[M]$. Assume that $\hat{\rho}(0)>0$, $\mathbf{H_0}$ and $\mathbf{H_2}$ are satisfied. Suppose that for all $k\in[M]$,
\begin{equation}
1-\frac{\underline{\ell}_k^2}{2(\underline{\ell}^2_k+|\mathbf{Re}(\mathfrak{l}_{k,\bar{n}})|\bar{\ell}_k)}< \sqrt{\frac{\eta_k \gamma_k}{\hat{\eta}_k \hat{\gamma}_k}}\leq 1,
\label{Eq:ConditionInstability_G1}
\end{equation}
or 
\begin{equation}
\begin{cases}
1+ \mathbf{l}_{k,\bar{n}}\leq \sqrt{\frac{\eta_k \gamma_k}{\hat{\eta}_k \hat{\gamma}_k}},& \textrm{if }\underline{\ell}^2_k>|\mathbf{Re}\{\mathfrak{l}_{k,\bar{n}}\}|\bar{\ell}_k,\\
1\leq \sqrt{\frac{\eta_k \gamma_k}{\hat{\eta}_k \hat{\gamma}_k}},& \textrm{if }\underline{\ell}^2_k=|\mathbf{Re}\{\mathfrak{l}_{k,\bar{n}}\}|\bar{\ell}_k,\\
1\leq \sqrt{\frac{\eta_k \gamma_k}{\hat{\eta}_k \hat{\gamma}_k}} \leq 1-\mathbf{l}_{k,\bar{n}},& \textrm{if }\underline{\ell}^2_k<|\mathbf{Re}\{\mathfrak{l}_{k,\bar{n}}\}|\bar{\ell}_k.
\end{cases}
\label{Eq:ConditionInstability_G2}
\end{equation}
where $\mathbf{l}_{k,\bar{n}}:=\underline{\ell}_k^2/2(\underline{\ell}^2_k-|\mathbf{Re}\{\mathfrak{l}_{k,\bar{n}}\}|\bar{\ell}_k)$.
Then, there exists $\lambda>0$ such that, for all initial condition $(\rho(0),\hat{\rho}(0))\in\mathbf{B}_{\lambda}(\mathcal{H}_{n}\times\mathcal{H}_{\bar{n}})\bigcap \mathcal{S}\times\partial \mathcal{S}$ with $n\neq\bar{n}$, the solutions $(\rho(t),\hat{\rho}(t))$ of the coupled system~\eqref{Eq:SME_W}--\eqref{Eq:SME_filter_W} exit $\mathbf{B}_{\lambda}(\mathcal{H}_{n}\times\mathcal{H}_{\bar{n}})$ in finite time almost surely. 
\label{Lemma:Unstability_General}
\end{lemma}
\proof
To prove the instability of $\mathcal{I}(\mathcal{H}_n)\times\mathcal{I}(\mathcal{H}_{\bar{n}})$ with $n\neq \bar{n}$, we can apply \hyperref[Prop:Instability]{Proposition~\ref*{Prop:Instability}} with the Lyapunov function $-\log \mathrm{Tr}(\hat{\rho}P_n)\geq 0$. By \hyperref[Lemma:PosDef invariant]{Lemma~\ref*{Lemma:PosDef invariant}}, $\hat{\rho}(0)>0$ implies $\hat{\rho}(t)>0$ for all $t\geq0$ almost surely, and therefore $\mathrm{Tr}(\hat{\rho}(t)P_n)>0$ for all $t\geq0$. The infinitesimal generator of the Lyapunov function is given by
\begin{equation*}
\mathscr{L}  -\log \mathrm{Tr}(\hat{\rho}P_n) =-\frac{\mathscr{L} \mathrm{Tr}(\hat{\rho}P_n)}{\mathrm{Tr}(\hat{\rho}P_n)}+\frac{\sum^m_{k=1}\big(\hat{\mathsf{G}}^{n}_{k}(\hat{\rho})\big)^2}{2\mathrm{Tr}(\hat{\rho}P_n)^2},
\end{equation*}
where $\hat{\mathsf{G}}^{n}_{k}(\hat{\rho})=\sqrt{\hat{\eta}_k\hat{\gamma}_k}\Delta_{k,n}(\hat{\rho})\mathrm{Tr}(\hat{\rho}P_n)$ and
\begin{equation*}
\begin{split}
\mathscr{L} \mathrm{Tr}(\hat{\rho}P_n)=&-u\Theta_{\bar{n}}(\hat{\rho})+\sum^m_{k=1}\sqrt{\hat{\eta}_k\hat{\gamma}_k}\Delta_{k,n}(\hat{\rho})\mathcal{T}_k(\rho,\hat{\rho})\mathrm{Tr}(\hat{\rho}P_{n}).
\end{split}
\end{equation*}
Due to \textbf{H2}, for all $(\rho,\hat{\rho})\in\mathbf{B}_r(\mathcal{H}_n\times\mathcal{H}_{\bar{n}})$ with $r\in(0,\epsilon]$ where $\epsilon$ is defined in $\mathbf{H_2}$, we have
\begin{equation*}
\frac{\mathscr{L} \mathrm{Tr}(\hat{\rho}P_n)}{\mathrm{Tr}(\hat{\rho}P_n)}=\sum^m_{k=1}\sqrt{\hat{\eta}_k\hat{\gamma}_k}\Delta_{k,n}(\hat{\rho})\mathcal{T}_k(\rho,\hat{\rho}),
\end{equation*}
which implies 
\begin{align}
&\liminf_{(\rho,\hat{\rho})\rightarrow \mathcal{I}(\mathcal{H}_{n})\times \mathcal{I}(\mathcal{H}_{\bar{n}})}\frac{\mathscr{L} \mathrm{Tr}(\hat{\rho}P_n)}{\mathrm{Tr}(\hat{\rho}P_n)}\nonumber\\
&~~=4\sum^m_{k=1}\sqrt{\hat{\eta}_k\hat{\gamma}_k}\big(\mathbf{Re}\{\mathfrak{l}_{k,n}\}-\mathbf{Re}\{\mathfrak{l}_{k,\bar{n}}\}\big)\big( \sqrt{\eta_k\gamma_k}\mathbf{Re}\{\mathfrak{l}_{k,n}\}-\sqrt{\hat{\eta}_k\hat{\gamma}_k}\mathbf{Re}\{\mathfrak{l}_{k,\bar{n}}\} \big).\label{Eq:InstaGen_liminf}
\end{align}
Moreover, we have
\begin{align}
&\limsup_{(\rho,\hat{\rho})\rightarrow \mathcal{I}(\mathcal{H}_{n})\times \mathcal{I}(\mathcal{H}_{\bar{n}})}\frac{\sum^m_{k=1}\big(\hat{\mathsf{G}}^{n}_{k}(\hat{\rho})\big)^2}{2\mathrm{Tr}(\hat{\rho}P_n)^2}\leq 2\sum^{m}_{k=1} \hat{\eta}_k \hat{\gamma}_k \big(\mathbf{Re}\{\mathfrak{l}_{k,n}\}-\mathbf{Re}\{\mathfrak{l}_{k,\bar{n}}\}\big)^2.\label{Eq:InstaGen_limsup}
\end{align}
The condition~\eqref{Eq:ConditionInstability_G1} or~\eqref{Eq:ConditionInstability_G2} guarantees that the $k$-th term of RHS of~\eqref{Eq:InstaGen_liminf} is strictly greater than the $k$-th term of RHS of~\eqref{Eq:InstaGen_limsup} for all $k\in[M]$ such that $\mathbf{Re}\{\mathfrak{l}_{k,n}\}\neq \mathbf{Re}\{\mathfrak{l}_{k,\bar{n}}\}$. Note that it is impossible that $\mathbf{Re}\{\mathfrak{l}_{k,n}\}=\mathbf{Re}\{\mathfrak{l}_{k,\bar{n}}\}$ for all $k\in[M]$ due to the orthogonality of the projections $\{P_n\}^M_{n=1}$.
Therefore, there exist $C>0$ and a sufficiently small constant $\lambda\in(0,r)$ such that, for all $n\neq \bar{n}$ 
\begin{equation}
\mathscr{L}-\log \mathrm{Tr}(\hat{\rho}P_n) \leq -C,~\forall (\rho,\hat{\rho})\in\mathbf{B}_{\lambda}(\mathcal{H}_n\times\mathcal{H}_{\bar{n}}).
\end{equation}
Therefore, by applying \hyperref[Prop:Instability]{Proposition~\ref*{Prop:Instability}}, for any $(\rho(0),\hat{\rho}(0))\in\mathbf{B}_{\lambda}(\mathcal{H}_{n}\times\mathcal{H}_{\bar{n}})\bigcap \mathcal{S}\times\partial \mathcal{S}$ with $n\neq\bar{n}$, $\mathbb{E}(\tau^{n,\bar{n}}_{\lambda})<-\frac{1}{C}\log \mathrm{Tr}(\hat{\rho}(0)P_n)<\infty$. Then, by Markov inequality, we have $\mathbb{P}(\tau^{n,\bar{n}}_{\lambda}<\infty)=1$.
\hfill$\square$
\begin{remark}
\hyperref[Lemma:Unstability_Special]{Lemma~\ref*{Lemma:Unstability_Special}} and \hyperref[Lemma:Unstability_General]{Lemma~\ref*{Lemma:Unstability_General}} are analogous to~\cite[Section 4.2.1]{liang2021robustness}. The range of the model parameters $\hat{\eta}_k$ and $\hat{\gamma}_k$ provided in \hyperref[Lemma:Unstability_Special]{Lemma~\ref*{Lemma:Unstability_Special}} and \hyperref[Lemma:Unstability_General]{Lemma~\ref*{Lemma:Unstability_General}} ensuring the instability is not the optimal one.
\end{remark}
\subsubsection{Recurrence properties of the coupled trajectories} 
Based on the support theorem (\hyperref[Thm:Support]{Theorem~\ref*{Thm:Support}}), the deterministic control systems corresponding to the Stratonovich form of~\eqref{Eq:SME_W}--\eqref{Eq:SME_filter_W} is given by
\begin{align}
&\dot{\rho}_{v}(t)=\tilde{\mathcal{L}}^u_{\omega,\gamma,\eta}(\rho_{v}(t))+\sum^{m}_{k=1}\mathcal{G}^k_{\eta_k,\gamma_k}(\rho_{v}(t))V_k(t),\label{Eq:ODE}\\
&\dot{\hat{\rho}}_{v}(t)=\tilde{\mathcal{L}}^u_{\hat{\omega},\hat{\gamma},\hat{\eta}}(\hat{\rho}_{v}(t))+\sum^{m}_{k=1}\mathcal{G}^k_{\hat{\eta}_k,\hat{\gamma}_k}(\hat{\rho}_{v}(t))V_k(t),\label{Eq:ODE_F}
\end{align}
where $\rho_v(0)=\rho(0)$, $\hat{\rho}_v(0)=\hat{\rho}(0)$, $V_k(t):=v_k(t)+\sqrt{\eta_k \gamma_k}\mathrm{Tr}((L_k+L_k^*)\rho_v(t))$ where $v_k(t)\in\mathcal{V}$ is the bounded control input. Here
\begin{equation*}
\begin{split}
\tilde{\mathcal{L}}^u_{\omega,\gamma,\eta}(\rho):=&-i[\omega H_0+u H_1,\rho]\\
&+\sum^m_{k=1}\frac{\gamma_k}{2}\Big(2(1-\eta_k)L_k \rho L_k^*-(L^*_kL_k+\eta_k L_k^2)\rho\\
&-\rho(L^*_kL_k+\eta_k {L_k^*}^2)+\eta_k\mathrm{Tr}\big((L_k+L^*_k)^2\rho\big)\rho  \Big),
\end{split}
\end{equation*}
and $\mathcal{G}^k_{\eta_k,\gamma_k}(\rho)$ is defined as in~\eqref{Eq:SME}. By the support theorem (\hyperref[Thm:Support]{Theorem~\ref*{Thm:Support}}), the set $\mathcal{S}$ is invariant for~\eqref{Eq:ODE} and~\eqref{Eq:ODE_F}. 

For $l\in\mathbb{Z}$ and $\xi\in \mathbb{R}^N$, define
\begin{equation*}
\begin{split}
\mathbf{M}_{l,\xi}:=[\xi, H_1\xi, &L_1^*H_1\xi,\dots,L_m^*H_1\xi,\dots,H_1^l\xi,L_1^*H_1^l\xi,\dots,L_m^*H_1^l\xi].    
\end{split}
\end{equation*} 
Then, we introduce the following assumption on the control Hamiltonian and the measurement operators.
\smallskip
\begin{itemize}
\item[\textbf{H3}:] 
$\exists l\in\mathbb{Z}$ such that $\mathrm{rank}(\mathbf{M}_{l,\xi})=N$ for all $\xi\in\mathscr{B}$.
\end{itemize}
\smallskip
Inspired by~\cite[Lemma 6.1]{liang2021robustness},~\cite[Lemma 9]{liang2022GHZ} and~\cite[Theorem 4.7]{baxendale1991invariant}, in the following lemmas, we analyze the possibility of constructing trajectories of~\eqref{Eq:ODE}--\eqref{Eq:ODE_F} which enter an arbitrarily small neighbourhood of the target subspace. 
Before stating the results, for $k\in[M]$ and $n\in[M]$, we define $\Lambda_{k,n}(\rho):=\mathrm{Tr}((L_k+L_k^*)^2{\rho})-4|\mathbf{Re}\{\mathfrak{l}_{k,n}\}|^2$, $\boldsymbol\Delta_{k,n}:=\{\rho\in\mathcal{S}|\,\Delta_{k,n}(\rho)=0\}$, and the ``variance function'' $\mathscr{V}_k(\rho):=\mathrm{Tr}((L_k+L_k^*)^2\rho)-\mathrm{Tr}((L_k+L_k^*)\rho)^2$ of $L_k+L_k^*$.
\begin{lemma}
Assume that \emph{\textbf{H0}} and \emph{\textbf{H3}} hold true. In addition, suppose that 
\begin{enumerate}
\item[(i)] $\overline{C}_{k,\bar{n}}\leq 0$ or $\underline{C}_{k,\bar{n}}\geq 0$ for each $k\in[M]$,
\item[(ii)] for any $\hat{\rho}(0)\in\{\hat\rho\in\mathcal{S}| \, \mathrm{Tr}(\hat{\rho}P_{\bar{n}})=0\},$  there exists a control $v(t)\in\mathcal{V}^m$ such that for all $t\in(0,\delta),$ with $\delta>0$ sufficiently small, $u\big(\hat{\rho}_{v}(t)\big)\neq 0$, for some solution $\hat{\rho}_{v}(t)$ of~\eqref{Eq:ODE_F}.
\end{enumerate}
Then, for all $\varepsilon>0$ and $(\rho(0),\hat{\rho}(0))$ belongs to $$\left(\mathcal{S}\times\mathcal{S}\right)\setminus\big\{\mathbf{B}_{\varepsilon}(\mathcal{H}_{\bar{n}}\times\mathcal{H}_{\bar{n}})\cup\bigcup_{n\neq \bar{n}}\mathcal{I}(\mathcal{H}_{n})\times\mathcal{I}(\mathcal{H}_{\bar{n}})\big\},$$ there exist $T\in(0,\infty)$ and $v(t)\in\mathcal{V}^m$ such that the trajectory $(\rho_v(t),\hat{\rho}_{v}(t))$ of the coupled deterministic system~\eqref{Eq:ODE}--\eqref{Eq:ODE_F} enters $\mathbf{B}_{\varepsilon}(\mathcal{H}_{\bar{n}}\times\mathcal{H}_{\bar{n}})$ for $t<T$.
\label{Lemma:Reachability ODE - 1}
\end{lemma}
\proof
From~\eqref{Eq:ODE}--\eqref{Eq:ODE_F} we have
\begin{align}
&\frac{d}{dt}\mathrm{Tr}(\rho_v(t)P_{\bar{n}})=-u\Theta_{\bar{n}}(\rho_v(t))+\frac{\mathrm{Tr}(\rho_v(t)P_{\bar{n}})}{2}\sum^m_{k=1}\eta_k\gamma_k\Lambda_{k,\bar{n}}(\rho_v(t))\nonumber\\
&~~~~~~~~~~~~~~~~~~~~~~+\mathrm{Tr}(\rho_v(t)P_{\bar{n}})\sum^m_{k=1}\sqrt{\eta_k\gamma_k}\Delta_{k,\bar{n}}(\rho_v(t))V_k(t),\label{Eq:ODE_Target}\\
&\frac{d}{dt}\mathrm{Tr}(\hat{\rho}_v(t)P_{\bar{n}})=-u\Theta_{\bar{n}}(\hat{\rho}_v(t))+\frac{\mathrm{Tr}(\hat{\rho}_v(t)P_{\bar{n}})}{2}\sum^m_{k=1}\hat{\eta}_k\hat{\gamma}_k\Lambda_{k,\bar{n}}(\hat{\rho}_v(t))\nonumber\\
&~~~~~~~~~~~~~~~~~~~~~~+\mathrm{Tr}(\hat{\rho}_v(t)P_{\bar{n}})\sum^m_{k=1}\sqrt{\hat{\eta}_k\hat{\gamma}_k}\Delta_{k,\bar{n}}(\hat{\rho}_v(t))V_k(t),\label{Eq:ODE_F_Target}
\end{align}
where $\rho_v(0)=\rho(0)$, $\hat{\rho}_v(0)=\hat{\rho}(0)$.
If $\mathrm{Tr}(\rho(0)P_{\bar{n}})=\mathrm{Tr}(\hat{\rho}(0)P_{\bar{n}})=0$, by following the arguments of \hyperref[Prop:ExitBoundary]{
Proposition~\ref*{Prop:ExitBoundary}}, one can show the existence of $v\in\mathcal{V}^m$ and $\delta>0$ such that $\mathrm{Tr}(\rho_v(0)P_{\bar{n}})>0$ and $\mathrm{Tr}(\hat{\rho}_v(0)P_{\bar{n}})>0$ for $t\in(0,\delta)$. Thus, without loss of generality, we suppose $\mathrm{Tr}(\rho(0)P_{\bar{n}})>0$ and $\mathrm{Tr}(\hat{\rho}(0)P_{\bar{n}})>0.$ Moreover, due to the inequality~\eqref{Eq:IneqLambda}, for each $k\in[M]$, if $\overline{C}_{k,\bar{n}}\leq 0$ or $\underline{C}_{k,\bar{n}}\geq 0$, then $\boldsymbol\Delta_{k,\bar{n}}=\mathcal{I}(\mathcal{H}_{\bar{n}})$. The proof can be concluded by applying the similar arguments as in the proof of~\cite[Lemma 6.1]{liang2021robustness}.
\hfill$\square$

\begin{lemma}
Assume that \emph{\textbf{H0}}, \emph{\textbf{H2}} and \emph{\textbf{H3}} are satisfied. 
In addition, suppose that 
\begin{enumerate}
\item[\emph{(i)}]  for any $k\in[M]$,
\begin{equation}
\left\{
\begin{split}
\sqrt{\eta_k \gamma_k} \,\underline{\mathbf{R}}_{k,\bar{n}}<\sqrt{\hat{\eta}_k \hat{\gamma}_k}\,\overline{\mathbf{R}}_{k,\bar{n}},\\
\sqrt{\hat{\eta}_k \hat{\gamma}_k}\,\underline{\mathbf{R}}_{k,\bar{n}}<\sqrt{\eta_k \gamma_k}\,\overline{\mathbf{R}}_{k,\bar{n}};
\end{split}
\right.
\label{Eq:ConditionParameterG}
\end{equation}
with $\underline{\mathbf{R}}_{k,\bar{n}}:=2\mathbf{Re}\{\mathfrak{l}_{k,\bar{n}}\}-\underline{\ell}_{k}$ and $\overline{\mathbf{R}}_{k,\bar{n}}:=2\mathbf{Re}\{\mathfrak{l}_{k,\bar{n}}\}+\underline{\ell}_{k}$.
\item[\emph{(ii)}] for any $\hat{\rho}\in\bigcap^m_{k=1}\boldsymbol\Delta_{k,\bar{n}}\setminus \mathcal{I}(\mathcal{H}_{\bar{n}})$, 
\begin{equation}
u\Theta_{\bar{n}}(\hat{\rho})<\frac{\mathrm{Tr}(\hat{\rho}P_{\bar{n}})}{2}\sum^m_{k=1}\hat{\eta}_k\hat{\gamma}_k\mathscr{V}_k(\hat{\rho}).
\label{Eq:Cond_uRho}
\end{equation}
\end{enumerate}
Then, for all $\varepsilon>0$ and any given initial state $(\rho(0),\hat{\rho}(0)) \in \mathcal{S}\times\mathrm{int}(\mathcal{S})\setminus\mathbf{B}_{\varepsilon}(\mathcal{H}_{\bar{n}}\times\mathcal{H}_{\bar{n}})$, there exist $T\in(0,\infty)$ and $v(t)\in\mathcal{V}^m$ such that the trajectory of~\eqref{Eq:ODE}--\eqref{Eq:ODE_F} enters $\mathbf{B}_{\varepsilon}(\mathcal{H}_{\bar{n}}\times\mathcal{H}_{\bar{n}})$ for $t<T$.
\label{Lemma:Reachability ODE - 2}
\end{lemma}
\textit{Sketch of the proof.}
The lemma can be proved in three steps by following the similar arguments as in the proof of~\cite[Lemma 4.8]{liang2021robustness} and~\cite[Lemma 9]{liang2022GHZ}:
\begin{enumerate}
\item $\forall \hat{\rho}(0)>0$, $\exists v\in\mathcal{V}^m$ s.t. $u(\hat{\rho}_v(t))\neq 0$ for some $t>0$.
\item $\exists T_1\in(0,\infty)$, $\exists v\in\mathcal{V}^m$ s.t. $\hat{\rho}_v(T_1)\in B_{\varepsilon}(\mathcal{H}_{\bar{n}})$.
\item $\forall \varepsilon\in(0,\epsilon)$, $\exists T_2\in(0,\infty)$ and $v(t)\in\mathcal{V}$ s.t. $(\rho_v(t),\hat{\rho}_v(t))$ enters $\mathbf{B}_{\varepsilon}(\mathcal{H}_{\bar{n}}\times\mathcal{H}_{\bar{n}})$ with $t<T_2$.
\end{enumerate}

Based on the reachability of the deterministic coupled systems~\eqref{Eq:ODE}--\eqref{Eq:ODE_F}, we can now state the following recurrence result for the stochastic coupled system~\eqref{Eq:SME_W}--\eqref{Eq:SME_filter_W}, which can be proved by the same arguments as in the proof of~\cite[Lemma 4.10]{liang2021robustness}.
\begin{lemma}
Assume that \emph{\textbf{H0}} and \emph{\textbf{H3}} hold true. If one of the following two conditions is satisfied
\begin{enumerate}
\item for $(\rho(0),\hat{\rho}(0))\in\mathcal{S}\times\mathcal{S}\setminus\bigcup_{n\neq\bar{n}}\mathcal{I}(\mathcal{H}_{n})\times\mathcal{I}(\mathcal{H}_{\bar{n}})$, \emph{\textbf{H1}} and the hypotheses in \hyperref[Lemma:Unstability_Special]{Lemma~\ref*{Lemma:Unstability_Special}} and \hyperref[Lemma:Reachability ODE - 1]{Lemma~\ref*{Lemma:Reachability ODE - 1}} hold true,
\item for $(\rho(0),\hat{\rho}(0))\in\mathcal{S}\times\mathrm{int}(\mathcal{S})$, \emph{\textbf{H2}} and the hypotheses in \hyperref[Lemma:Unstability_General]{Lemma~\ref*{Lemma:Unstability_General}} and \hyperref[Lemma:Reachability ODE - 2]{Lemma~\ref*{Lemma:Reachability ODE - 2}} hold true.
\end{enumerate}
Then, for all $\varepsilon>0$ one has 
$
\mathbb{P}(\tau^{\bar{n},\bar{n}}_{\varepsilon} < \infty)=1.
$
\label{Lemma:Reachability}
\end{lemma}

\subsection{Application of Theorem~\ref{Thm:Exp Stab General}: sufficient conditions ensuring exponential stabilization}
\label{sec:explicit}
In this section, we establish conditions on the feedback controller $u(\hat{\rho})$, the control Hamiltonian, and the domain of the model parameters, which ensure almost sure exponential stabilization of the coupled system~\eqref{Eq:SME_W}--\eqref{Eq:SME_filter_W} toward the target subspace $\mathcal{H}_{\bar{n}}\times\mathcal{H}_{\bar{n}}$. 
Define 
$$
K:=\frac{1}{2(M-1)}\min\Big\{\min_{k\in[m]}\{\eta_k\gamma_k\underline{\ell}_k^2\},\min_{k\in[m]}\{\hat{\eta}_k\hat{\gamma}_k\underline{\ell}_k^2\}\Big\}>0
$$ and 
\begin{equation*}
\begin{split}
C_{\bar{n}}:=\frac{1}{2}\min\Big\{ \min_{k\in[M]}\{\eta_k\gamma_k\underline{\ell}_{k}\},&\min_{k\in[M]}\{\hat{\eta}_k\hat{\gamma}_k\underline{\ell}_{k}\}\\
&-4\sum^m_{k=1}\bar{\ell}_k|\mathbf{Re}\{\mathfrak{l}_{k,\bar{n}}\}|\sqrt{\hat{\eta}_k\hat{\gamma}_k}|\sqrt{\eta_k\gamma_k}-\sqrt{\hat{\eta}_k\hat{\gamma}_k}|\Big\}.
\end{split}
\end{equation*}
\begin{theorem}
Consider the coupled system~\eqref{Eq:SME_W}--\eqref{Eq:SME_filter_W}. Suppose that the assumptions given in \hyperref[Lemma:Reachability]{Lemma~\ref*{Lemma:Reachability}} and 
\hyperref[Cor:ExitBoundaryLemma]{Corollary~\ref*{Cor:ExitBoundary}} are satisfied, and 
\begin{align}
&\min_{k\in[M]}\{\hat{\eta}_k\hat{\gamma}_k\underline{\ell}_{k}\}>4\sum^m_{k=1}\bar{\ell}_k|\mathbf{Re}\{\mathfrak{l}_{k,\bar{n}}\}|\sqrt{\hat{\eta}_k\hat{\gamma}_k}|\sqrt{\eta_k\gamma_k}-\sqrt{\hat{\eta}_k\hat{\gamma}_k}|.
\label{Eq:ConditionParameter}
\end{align}
Then, $\mathcal{H}_{\bar{n}}\times\mathcal{H}_{\bar{n}}$ is almost surely exponentially stable with sample Lyapunov exponent less than or equal to $-C_{\bar{n}}-K$ if $\overline{C}_{k,\bar{n}}\leq 0$ or $\underline{C}_{k,\bar{n}}\geq 0$ for each $k\in[M]$, and $-C_{\bar{n}}$ otherwise.
\label{Thm:ExpStab}
\end{theorem}
\proof
Define $V_{\bar{n}}(\rho):=\sum_{n\neq\bar{n}}\sqrt{\mathrm{Tr}(\rho P_n)}$. Consider the candidate Lyapunov function
$
\mathbf{V}_{\bar{n}}(\rho,\hat{\rho})=V_{\bar{n}}(\rho)+V_{\bar{n}}(\hat{\rho}).
$
Due to \hyperref[Lemma:PosDef invariant]{Lemma~\ref*{Lemma:PosDef invariant}}, $\Gamma=\mathrm{int}(\mathcal{S}_N)\times\mathrm{int}(\mathcal{S})$ is almost surely invariant. Since \hyperref[Cor:ExitBoundary]{Corollary~\ref*{Cor:ExitBoundary}}, $\rho(t)$ and $\hat{\rho}(t)$ can enter in $\Gamma$ in finite time almost surely. $\mathbf{V}_{\bar{n}}(\rho,\hat{\rho})$ is continuous on $\mathcal{S}\times\mathcal{S}$ and twice continuously differentiable on $\Gamma.$

It is easy to verify that, for all $(\rho,\hat{\rho})\in\mathcal{S}\times\mathcal{S}$, $\mathbf{V}_{\bar{n}}(\rho,\hat{\rho})\geq E_{\bar{n}}(\rho)+E_{\bar{n}}(\hat{\rho})$. Note that for some $k\in[M]$, there may exist $n\in[M]$ such that $\mathbf{Re}\{\mathfrak{l}_{k,n}\} =  \mathbf{Re}\{\mathfrak{l}_{k,\bar{n}}\}$. For each $k\in[M]$, we define the following sets
\begin{equation*}
\begin{split}
&S^{\neq}_{k,\bar{n}}:=\{n\in[M]\setminus\{\bar{n}\}|\,\mathbf{Re}\{\mathfrak{l}_{k,n}\} \neq  \mathbf{Re}\{\mathfrak{l}_{k,\bar{n}}\}\},\\
&S^{=}_{k,\bar{n}}:=\{n\in[M]\setminus\{\bar{n}\}|\,\mathbf{Re}\{\mathfrak{l}_{k,n}\} =  \mathbf{Re}\{\mathfrak{l}_{k,\bar{n}}\}\}.
\end{split}
\end{equation*}
Due to \textbf{H0} and the orthogonality of $\{P_n\}^M_{n=1}$, we have $\bigcup_{k}S^{\neq}_{k,\bar{n}}=[M]\setminus\{\bar{n}\}$. Note that $S^{=}_{k,\bar{n}}$ may be empty for some $k$.
We recall $\Delta_{k,n}(\rho)=2\mathbf{Re}\{\mathfrak{l}_{k,n}\}-\mathrm{Tr}((L_k+L_k^*)\rho).$ By a straightforward calculation, for all $n\in[M]$ and $k\in[M]$, $ |\Delta_{k,n}(\rho)|\leq 2\bar{\ell}_k+|\Delta_{k,\bar{n}}(\rho)|$ and there exists a positive constant $c_n>0$ such that $|\Theta_{n}(\rho)|\leq c_n \sqrt{\mathrm{Tr}(\rho P_n)}$. In a sufficiently small neighbourhood of target subspace, $|u|\leq c V_{\bar{n}}(\hat{\rho})^{\alpha}$ with $\alpha>1$ and $c>0$ under \textbf{H1} and $c=0$ under \textbf{H2}, for each $k$ and $n\in S^{\neq}_{k,\bar{n}}$, $|\Delta_{k,n}(\rho)|\geq 2\underline{\ell}_k-|\Delta_{k,\bar{n}}(\rho)|>0$. 
The infinitesimal generator of $\mathbf{V}_{\bar{n}}(\rho,\hat{\rho})$ is given by
\begin{equation*}
\begin{split}
\mathscr{L}&\mathbf{V}_{\bar{n}}(\rho,\hat{\rho})\\
=&-\frac{u}{2}\sum_{n\neq \bar{n}}\Big(\frac{\Theta_{n}(\rho)}{\sqrt{\mathrm{Tr}(\rho P_n)}}+\frac{\Theta_{n}(\hat{\rho})}{\sqrt{\mathrm{Tr}(\hat{\rho} P_n)}}\Big)-\frac{1}{8}\sum^m_{k=1}\eta_k\gamma_k\sum_{n\neq \bar{n}}\Delta_{k,n}(\rho)^2\sqrt{\mathrm{Tr}(\rho P_n)}\\
&-\frac{1}{8}\sum^m_{k=1}\hat{\eta}_k\hat{\gamma}_k\sum_{n\neq \bar{n}}\Delta_{k,n}(\hat{\rho})^2\sqrt{\mathrm{Tr}(\hat{\rho} P_n)}+\frac{1}{2}\sum^m_{k=1}\sqrt{\hat{\eta}_k\hat{\gamma}_k}\mathcal{T}_k(\rho,\hat{\rho})\sum_{n\neq \bar{n}}\Delta_{k,n}(\hat{\rho})\sqrt{\mathrm{Tr}(\hat{\rho}P_n)}\\
\leq&|u|\sum_{n\neq \bar{n}}c_n-\frac{1}{8}\sum^m_{k=1}\eta_k\gamma_k(2\underline{\ell}_k-|\Delta_{k,\bar{n}}(\rho)|)^2\sum_{n\in S^{\neq}_{k,\bar{n}}}\sqrt{\mathrm{Tr}(\rho P_n)}\\
&-\frac{1}{8}\sum^m_{k=1}\hat{\eta}_k\hat{\gamma}_k(2\underline{\ell}_k-|\Delta_{k,\bar{n}}(\hat{\rho})|)^2\sum_{n\in S^{\neq}_{k,\bar{n}}}\sqrt{\mathrm{Tr}(\hat{\rho} P_n)}+\frac{V_{\bar{n}}(\hat{\rho})}{2}\sum^m_{k=1}\sqrt{\hat{\eta}_k\hat{\gamma}_k}|\mathcal{T}_k(\rho,\hat{\rho})|(2\bar{\ell}_{k}+|\Delta_{k,\bar{n}}(\hat{\rho})|)\\
\leq & -\frac{1}{8}\min_{k\in[M]}\{\eta_k\gamma_k(2\underline{\ell}_k-|\Delta_{k,\bar{n}}(\rho)|)^2\} V_{\bar{n}}(\rho)+c \sum_{n\neq \bar{n}}c_n V_{\bar{n}}(\hat{\rho})^\alpha\\
&-\frac{1}{8}\min_{k\in[M]}\{\hat{\eta}_k\hat{\gamma}_k(2\underline{\ell}_k-|\Delta_{k,\bar{n}}(\hat{\rho})|)^2\} V_{\bar{n}}(\hat{\rho})+\frac{V_{\bar{n}}(\hat{\rho})}{2}\sum^m_{k=1}\sqrt{\hat{\eta}_k\hat{\gamma}_k}|\mathcal{T}_k(\rho,\hat{\rho})|(2\bar{\ell}_{k}+|\Delta_{k,\bar{n}}(\hat{\rho})|)\\
\leq& -\mathbf{C}_{\bar{n}}(\rho,\hat{\rho})\mathbf{V}_{\bar{n}}(\rho,\hat{\rho}),
\end{split}
\end{equation*}
where 
\begin{equation*}
\begin{split}
\mathbf{C}_{\bar{n}}(\rho,\hat{\rho}):=\min\Big\{&\frac{1}{8}\min_{k\in[M]}\{\eta_k\gamma_k(2\underline{\ell}_k-|\Delta_{k,\bar{n}}(\rho)|)^2\}, \frac{1}{8}\min_{k\in[M]}\{\hat{\eta}_k\hat{\gamma}_k(2\underline{\ell}_k-|\Delta_{k,\bar{n}}(\hat{\rho})|)^2\}\\
&-c \sum_{n\neq \bar{n}}c_n V_{\bar{n}}(\hat{\rho})^{\alpha-1}-\frac{1}{2}\sum^m_{k=1}\sqrt{\hat{\eta}_k\hat{\gamma}_k}|\mathcal{T}_k(\rho,\hat{\rho})|(2\bar{\ell}_{k}+|\Delta_{k,\bar{n}}(\hat{\rho})|)\Big\}.
\end{split}
\end{equation*}
Thus, we have
\begin{equation*}
\begin{split}
&\limsup_{(\rho,\hat{\rho})\rightarrow\mathcal{I}(\mathcal{H}_{\bar{n}})\times\mathcal{I}(\mathcal{H}_{\bar{n}})}\frac{\mathscr{L}\mathbf{V}_{\bar{n}}(\rho,\hat{\rho})}{\mathbf{V}_{\bar{n}}(\rho,\hat{\rho})}\leq \limsup_{(\rho,\hat{\rho})\rightarrow\mathcal{I}(\mathcal{H}_{\bar{n}})\times\mathcal{I}(\mathcal{H}_{\bar{n}})}-\mathbf{C}_{\bar{n}}(\rho,\hat{\rho})\leq -C_{\bar{n}}<0,
\end{split}
\end{equation*}
where the positivity of $C_{\bar{n}}$ is guaranteed by~\eqref{Eq:ConditionParameter}.

Furthermore, for all $k\in[m]$, if $\overline{C}_{k,\bar{n}}\leq 0$, then $\Delta_{k,\bar{n}}(\rho)\leq 0$ and $\Delta_{k,n}(\rho)\geq 0$ with $n\neq \bar{n}$ in a sufficiently small neighbourhood of the target subset. Also if $\underline{C}_{k,\bar{n}}\geq 0$, then $\Delta_{k,\bar{n}}(\rho)\geq 0$ and $\Delta_{k,n}(\rho)\leq 0$ with $n\neq \bar{n}$ in a sufficiently small neighbourhood of the target subset. Thus, in such a neighbourhood, this implies $\Delta_{k,n}(\rho)^2=4|\mathbf{Re}\{\mathfrak{l}_{k,n}\} - \mathbf{Re}\{\mathfrak{l}_{k,\bar{n}}\}|^2+|\Delta_{k,\bar{n}}(\rho)|^2-4|\mathbf{Re}\{\mathfrak{l}_{k,n}\} - \mathbf{Re}\{\mathfrak{l}_{k,\bar{n}}\}||\Delta_{k,\bar{n}}(\rho)|$ for $n\neq \bar{n}$.
Then, in this neighbourhood, by a straightforward calculation, we have
\begin{equation*}
\begin{split}
\sum^m_{k=1}&g_k(\rho,\hat{\rho})^2\\
=&\frac{1}{4\mathbf{V}_{\bar{n}}(\rho,\hat{\rho})}\sum^m_{k=1}\Big(\sqrt{\eta_k\gamma_k}\Big(\sum_{n\neq \bar{n}}\sqrt{\mathrm{Tr}(\rho P_n)}  \Delta_{k,n}(\rho)\Big)+\sqrt{\hat{\eta}_k\hat{\gamma}_k}\Big(\sum_{n\neq \bar{n}}\sqrt{\mathrm{Tr}(\hat{\rho} P_n)} \Delta_{k,n}(\hat{\rho})\Big)  \Big)^2\\
\geq&\frac{1}{M-1}\min\Big\{\min_{k\in[m]}\{\eta_k\gamma_k\underline{\ell}_k^2\}-X_{\bar{n}}(\rho),\min_{k\in[m]}\{\hat{\eta}_k\hat{\gamma}_k\underline{\ell}_k^2\}-X_{\bar{n}}(\hat{\rho})\Big\},
\end{split}
\end{equation*}
where $X_{\bar{n}}(\rho):=\frac{1}{4}\sum^m_{k=1}|\Delta_{k,\bar{n}}(\rho)|(|\Delta_{k,\bar{n}}(\rho)|-4\bar{\ell}_k)$. 
Thus, we have
\begin{equation*}
\liminf_{(\rho,\hat{\rho})\rightarrow\mathcal{I}(\mathcal{H}_{\bar{n}})\times\mathcal{I}(\mathcal{H}_{\bar{n}})}\frac{1}{2}\sum^m_{k=1}g_k(\rho,\hat{\rho})^2\geq K. 
\end{equation*}
Then, we can apply \hyperref[Thm:Exp Stab General]{Theorem~\ref*{Thm:Exp Stab General}} to conclude the proof.

\begin{remark}
In the above theorem, we do not optimize the estimation of the Lyapunov exponent and the range of the estimated model parameters. They could be improved by applying \hyperref[Thm:Exp Stab General]{Theorem~\ref*{Thm:Exp Stab General}} on the different Lyapunov function, see~\cite[Section 4.3]{liang2021robustness}.
\end{remark}

As an example of application of the previous results, we design following parametrized feedback laws. Define
\begin{equation}
u_{\bar{n}}(\hat{\rho})  = \alpha \big(1-\mathrm{Tr}(\hat{\rho} P_{\bar{n}})\big)^{\beta},
\label{Eq:u_Special}
\end{equation}
with $\alpha>0$ and $\beta \geq1$, then \textbf{H1} holds true. In order to construct a feedback controller satisfying \textbf{H2}, we define a continuously differentiable function $f:[0,1]\to[0,1]$,
\begin{equation*}
f(x) := 
\begin{cases}
0,&\text{if }x\in[0,\epsilon_1);\\
\frac12\sin\left(\frac{\pi(2x-\epsilon_1-\epsilon_2)}{2(\epsilon_2-\epsilon_1)}\right)+\frac12,&\text{if }x\in[\epsilon_1,\epsilon_2);\\
1,&\text{if }x\in(\epsilon_2,1],
\end{cases}
\end{equation*}
where $0<\epsilon_1<\epsilon_2<1$. Define 
\begin{equation}
u_{\bar{n}}(\hat{\rho})=\alpha \big(1-\mathrm{Tr}(\hat{\rho} P_{\bar{n}})\big)^{\beta}f(1-\mathrm{Tr}(\hat{\rho}P_{\bar{n}})),
\label{Eq:u_General1}
\end{equation}
with $\alpha>0$ and $\beta \geq1$, then \textbf{H2} holds true. Define
\begin{equation}
u_{\bar{n}}(\hat{\rho})=f(1-\mathrm{Tr}(\hat{\rho}P_{\bar{n}}))\sum^m_{k=1}\alpha_k \big(\Delta_{k,\bar{n}}(\hat{\rho})\big)^{\beta_k},
\label{Eq:u_General}
\end{equation}
with $\alpha_k>0$ and $\beta_k\geq1$ for all $k\in[M]$, then \textbf{H2} and~\eqref{Eq:Cond_uRho} hold true.

\section{Stabilization by the simplified filter}
\label{sec:SimplifiedFeedback}
In this section, we show that the previous sections are particularly of interest to prove the exponential stabilization toward the target subspace by a feedback depending only on a simplified filter which will be discussed here.
In the previous section, we have studied the feedback stabilization of the open quantum system~\eqref{Eq:SME} toward the target subspace by evolving the full filter state in time~\eqref{Eq:SME_filter}. 

Inspired by the discussion in~\cite[Section 6]{cardona2020exponential}, we observe that $u(\hat{\rho})$ designed in~\eqref{Eq:u_Special}, \eqref{Eq:u_General1} and \eqref{Eq:u_General} depend almost only on the diagonal elements of the estimated state $\hat{\rho}$ in the non-demolition basis $\mathscr{B}$, i.e., $\mathrm{Tr}(\hat{\rho}P_n)$ for $n\in[M]$, which satisfy $\mathrm{Tr}(\hat{\rho}P_n)\in[0,1]$ and $\sum^M_{n=1}\mathrm{Tr}(\hat{\rho}P_n)=1$. From the proof in \hyperref[Lemma:Unstability_General]{Lemma~\ref*{Lemma:Unstability_General}}, the dynamics of $\mathrm{Tr}(\hat{\rho}P_n)$ is given by
\begin{equation*}
\begin{split}
 d \mathrm{Tr}(\hat{\rho}(t)P_n)=-u(\hat{\rho}(t))\Theta_{{n}}(\hat{\rho}(t))dt&+\sum^m_{k=1}\sqrt{\hat{\eta}_k\hat{\gamma}_k}\Delta_{k,n}(\hat{\rho}(t))\mathrm{Tr}(\hat{\rho}(t)P_n)\times\\
 &~~~~~~~~~~~~\Big(dY_k(t)-\sqrt{\hat{\eta}_k \hat{\gamma}_k}\mathrm{Tr}\big((L_k+L^*_k)\hat{\rho}(t)\big)dt\Big),
 \end{split}
\end{equation*}   
where
$\Theta_{n}(\hat\rho)=\mathrm{Tr}(i[H_1,\hat{\rho}]P_{n})$ and 
$\Delta_{k,n}(\hat{\rho})=2\big(\mathbf{Re}\{\mathfrak{l}_{k,n}\}-\mathrm{Tr}((L_k+L^*_k)\hat{\rho}).
$
Since all measurement operators are diagonal in $\mathscr{B}$, for all $k\in[M]$, 
\begin{equation*}
\mathrm{Tr}((L_k+L^*_k)\hat{\rho})=2\sum^{M}_{j=1}\mathbf{Re}\{\mathfrak{l}_{k,j}\}\mathrm{Tr}(\hat{\rho}P_j).
\end{equation*}
Thus, the dynamics of the diagonal elements are almost self-contained, the only term involving extra-diagonal coefficients $\Theta_{{n}}(\hat{\rho})$ is supported by the control term. If we set the extra-diagonal terms as zero, then we can obtain the following $M$-dimensional autonomous stochastic differential equation, 
\begin{align*}
d \hat{q}_n(t)=2\sum^m_{k=1}&\sqrt{\hat{\eta}_k\hat{\gamma}_k}\big( \mathbf{Re}\{\mathfrak{l}_{k,n}\}-\Lambda_k(\hat{q}(t))\big)\hat{q}_{n}(t)\Big(dY_k(t)-2\sqrt{\hat{\eta}_k \hat{\gamma}_k}\Lambda_k(\hat{q}(t))dt\Big).
\end{align*}
Since all standard basis vectors of $\mathbb{C}^M$ are equilibria, the above stochastic differential equation is a candidate estimator for the system~\eqref{Eq:SME} without control input (see Theorem~\ref{Thm:QSR} and Remark~\ref{Rem:QSR}). 

Based on the analysis in~\cite{liang2019exponential,liang2022GHZ}, the term containing feedback controller in the simplified filter is necessary, as it ensures  that the target subspace is the only equilibrium. Moreover, we need to guarantee that the state components of the simplified filter remain non-negative and their sum equal to one. Inspired by~\cite[Section 6]{cardona2020exponential}, we approximate the dynamics of $\mathrm{Tr}(\hat{\rho}P_n)$ for $n\in[M]$ by the following $M$-dimensional autonomous system of $\hat{q}=\{\hat{q}_n\}^M_{n=1}\in\mathbb{R}^M$,
\begin{align}
d \hat{q}_n(t)=u(\hat{q}(t))\sum^M_{j=1}\Gamma_{n,j}\hat{q}_j(t)dt&+2\sum^m_{k=1}\sqrt{\hat{\eta}_k\hat{\gamma}_k}\big( \mathbf{Re}\{\mathfrak{l}_{k,n}\}-\Lambda_k(\hat{q}(t))\big)\hat{q}_{n}(t)\times\nonumber\\
&~~~~\Big(dY_k(t)-2\sqrt{\hat{\eta}_k \hat{\gamma}_k}\Lambda_k(\hat{q}(t))dt\Big),
\label{Eq:ApproxSDE}    
\end{align}
where $\Lambda_k(\hat{q}):=\sum^M_{j=1}\mathbf{Re}\{\mathfrak{l}_{k,j}\}\hat{q}_{j}$, $Y_k(t)$ represents the observation process of $k$-th probe, $u\in\mathcal{C}^1(\mathbb{R}^M,\mathbb{R})$ and the matrix $\Gamma\in\mathbb{R}^{M\times M}$ such that $\sum^M_{i=1}\Gamma_{i,j}=0$ for all $j\in[M]$ and $\Gamma_{i,i}\neq 0$ for all $i\neq \bar{n}$.
Due to the Girsanov Theorem~\cite[Theorem A.45]{barchielli2009quantum}, $Y(t)$ is a standard Brownian motion under a new probability measure $\mathbb{Q}$ which is equivalent to $\mathbb{P}$.

Since the system~\eqref{Eq:ApproxSDE} does not satisfy the global Lipschitz condition and linear growth condition (see~\cite[Section 2.3]{mao2007stochastic}), the explosion may occur in a finite time. In order to make use of the system~\eqref{Eq:ApproxSDE} as a simplified filter in the feedback design for stabilization of the quantum system~\eqref{Eq:SME}, we first verify the existence and uniqueness of the global solution of~\eqref{Eq:ApproxSDE} from given initial states. Then we will employ the similar arguments as in the previous sections to ensure exponential stabilization toward the target subspace. Denote by $\tau_e(\hat{q}(0),\omega):\mathbb{R}^M\times \Omega\rightarrow [0,\infty]$ the explosion time from the initial state $\hat{q}(0)$. 
Through a slight modification of~\cite[Section 3]{dalal2007stochastic},~\hyperref[Lemma:NeverReach]{Lemma~\ref*{Lemma:NeverReach}} and~\cite[Lemma 4.3.2]{mao2007stochastic},
we show that $\hat{q}$ is a good replacement of $\{\mathrm{Tr}(\hat{\rho}P_n)\}^M_{n=1}$ as it keeps the invariance property of the set $\mathcal{D}_M:=\{\hat{q}\in(0,1)^M|\sum^M_{n=1}\hat{q}_n=1\}$ for the system~\eqref{Eq:ApproxSDE}.
\begin{lemma}
For all initial state $\hat{q}(0)\in\mathcal{D}_M$, the system~\eqref{Eq:ApproxSDE} has a unique global solution $\hat{q}(t)\in\mathcal{D}_M$ for all $t\geq 0$ almost surely.
\label{Lemma:Invariance_q}
\end{lemma}
\proof
It is easy to verify that the local Lipschitz condition holds for~\eqref{Eq:ApproxSDE}. By~\cite[Theorem 5.2.8]{mao2007stochastic}, for any $\hat{q}(0)\in\mathbb{R}^M$, the system~\eqref{Eq:ApproxSDE} has a local unique strong solution for all $t\in[0,\tau_e)$ almost surely. 
Moreover, we have
\begin{equation*}
\begin{split}
d&\sum^M_{n=1}\hat{q}_n(t)=2\sum^m_{k=1}\sqrt{\hat{\eta}_k\hat{\gamma}_k}\Lambda_k(\hat{q}(t))\Big(1-\sum^M_{n=1}\hat{q}_{n}(t)\Big)\Big(dY_k(t)-2\sqrt{\hat{\eta}_k \hat{\gamma}_k}\Lambda_k(\hat{q}(t))dt\Big),
\end{split}
\end{equation*}
which implies that, for all $\hat{q}(0)\in\mathcal{D}_M$, $\sum^M_{n=1}\hat{q}_n(t)=1$ till the explosion time. 
For all $\hat{q}(0)\in\mathcal{D}_M$, we take $k_0\geq 0$ sufficiently large so that $\hat{q}_n\in[1/k_0,k_0]$ for all $n\in[M]$. For each integer $k\geq k_0$ , we define the stopping time 
$$\tau_k:=\inf\{t\in[0,\tau_e|\,\hat{q}_n(t)\notin(1/k,k),~\forall\,n\in[M] \},
$$
which is an increasing sequence and $\tau_k$ converges to $\tau_e$ as $k$ tends to infinity. Consider the function
$
V(\hat{q})=\sum^M_{j=1}\big( \hat{q}_j+1-\log \hat{q}_j\big),
$
which belongs to $\mathcal{C}^2(\mathbb{R}^M_{>0},\mathbb{R}_{>0})$.
Suppose that $u$ is bounded. We can show that there exists a constant $C>0$ such that $\mathscr{L} V(\hat{q}) \leq C(1+V(\hat{q}))$ for all $\hat{q}\in\mathcal{D}_M$. By following the similar arguments as in~\cite[Theorem 1]{dalal2007stochastic}, one deduces that $\tau_{\infty}=\infty$ almost surely. Thus, for all initial state $\hat{q}(0)\in\mathcal{D}_M$, $\mathbb{Q}( \hat{q}_n(t)>0, \forall\, 0\leq t<\tau_e )=1$ for all $n\in[M]$. Then, employing the arguments in~\cite[Proposition 3.5]{mirrahimi2007stabilizing}, the boundness of $u$ can be removed.
\hfill$\square$

Next, we make the following assumptions on the new feedback controller $u(\hat{q})$ according to the system~\eqref{Eq:ApproxSDE}:
\smallskip
\begin{itemize}
\item[\textbf{H0'}:] 
$u\in\mathcal{C}^1(\mathcal{D}_M,\mathbb{R})$, $u(e_{\bar{n}})=0$ and $u(e_n)\neq0$ for all $n\neq\bar{n}$, where $\{e_n\}^M_{n=1}$ is a standard basis in $\mathbb{R}^M$.
\item[\textbf{H1'}:] 
$|u(\hat{q})|\leq c(1-\hat{q}_{\bar{n}})^\alpha$ with $\alpha>1/2$ and $c>0$.
\item[\textbf{H2'}:] 
$u(\hat{q})=0$ for all $\hat{\rho}\in \{\hat{q}\in\mathcal{D}_M|1-\hat{q}_n<\epsilon\}$ with $\epsilon>0$.
\item[\textbf{H4'}:] 
$\{\hat{q}\in\mathcal{D}_M|\,u(\hat{q})=0\}$ does not contain the complete integral curves of the drift and the vector fields of the deterministic system corresponding to~\eqref{Eq:ApproxSDE}.
\end{itemize}
\smallskip
By repeating the similar arguments as in the previous section, we obtain the following theorem concerning the exponential stabilization of~\eqref{Eq:SME} by simplified filter.
\begin{theorem}
Consider the coupled system~\eqref{Eq:SME}--\eqref{Eq:ApproxSDE} with initial states belonging to $\mathcal{S}\times \mathcal{D}_M$. Assume that \emph{\textbf{H0'}}, \emph{\textbf{H3}}, \emph{\textbf{H4'}} and the condition~\eqref{Eq:ConditionParameter} hold true and there exists at least one $k\in[M]$ such that $\eta_k,\hat{\eta}_k\in(0,1)$. Moreover, suppose that one of the following conditions is satisfied
\begin{enumerate}
\item[\emph{(i)}] $\overline{C}_{k,\bar{n}}\leq 0$ or $\underline{C}_{k,\bar{n}}\geq 0$ for each $k\in[M]$, $\overline{\mathcal{K}}^+_{\bar{n}}\cup \overline{\mathcal{K}}^-_{\bar{n}}=[M]$, Condition~\eqref{Eq:ConditionInstability_S1} and \emph{\textbf{H1'}} hold true;
\item[\emph{(ii)}] $\overline{C}_{k,\bar{n}}\leq 0$ or $\underline{C}_{k,\bar{n}}\geq 0$, Condition~\eqref{Eq:ConditionInstability_G1} or~\eqref{Eq:ConditionInstability_G2}, Condition~\eqref{Eq:ConditionParameterG} and \emph{\textbf{H2'}} hold true;
\item[\emph{(iii)}] Condition~\eqref{Eq:ConditionInstability_G1} or~\eqref{Eq:ConditionInstability_G2}, Condition~\eqref{Eq:ConditionParameterG} and \emph{\textbf{H2'}} hold true, and 
for any $\hat{q}\in \bigcap^m_{k=1}\{\hat{q}\in \mathcal{D}_M\setminus e_{\bar{n}}|\, \mathbf{Re}\{\mathfrak{l}_{k,\bar{n}}\}=\Lambda_k(\hat{q})\}$, 
\begin{equation}
    u\sum^M_{j=1}\Gamma_{\bar{n},j}\hat{q}_j<2\hat{q}_{\bar{n}}\sum^m_{k=1}\hat{\eta}_k\hat{\gamma}_k\mathsf{V}_k(\hat{q}),
    \label{Eq:Cond_uq}
\end{equation}
where $\mathsf{V}_k(\hat{q}):=\sum^M_{j=1}\mathbf{Re}\{\mathfrak{l}_{k,\bar{n}}\}^2\hat{q}_j-\Lambda_k(\hat{q})^2$.

\end{enumerate}
Then, $\mathcal{H}_{\bar{n}}\times e_{\bar{n}}$ is almost surely exponentially stable with sample Lyapunov exponent less than or equal to $-C_{\bar{n}}-K$ if $\overline{C}_{k,\bar{n}}\leq 0$ or $\underline{C}_{k,\bar{n}}\geq 0$ for each $k\in[M]$, and $-C_{\bar{n}}$ otherwise.
\label{Thm:ExpStab_q}
\end{theorem}

Based on the feedback controllers designed in~\eqref{Eq:u_Special}, \eqref{Eq:u_General1} and \eqref{Eq:u_General}, we provide the following examples of the simplified feedback controller. Define 
\begin{equation}
u_{\bar{n}}(\hat{q})  = \alpha (1-\hat{q}_{\bar{n}})^{\beta},
\label{Eq:u_Special_q}
\end{equation}
where $\alpha>0$ and $\beta \geq1$, then \textbf{H1'} holds true. Define 
\begin{equation}
u_{\bar{n}}(\hat{q})=\alpha (1-\hat{q}_{\bar{n}})^{\beta}f(1-\hat{q}_{\bar{n}}),
\label{Eq:u_General_q1}
\end{equation}
with $\alpha>0$ and $\beta \geq1$, then \textbf{H2'} holds true. Define 
\begin{equation}
u_{\bar{n}}(\hat{q})=f(1-\hat{q}_{\bar{n}})\sum^m_{k=1}\alpha_k \big(\mathbf{Re}\{\mathfrak{l}_{k,\bar{n}}\}-\Lambda_k(\hat{q}) \big)^{\beta_k},
\label{Eq:u_General_q}
\end{equation}
with $\alpha_k>0$ and $\beta_k\geq1$ for all $k\in[M]$, then \textbf{H2'} and~\eqref{Eq:Cond_uq} hold true.

\section{Simulation}
\label{sec:sim}
In this section, we illustrate our results by numerical simulations in the case of three-qubit systems proposed in~\cite{liang2022GHZ}. We take $L_1=\sigma_z\otimes \mathbf{I}\otimes \sigma_z+2 \sigma_z\otimes \sigma_z\otimes \mathbf{I}$ and $L_2 = \sigma_x\otimes \sigma_x\otimes \sigma_x$. Set $H_0=L_1$ and $H_1=(\mathbf{I}\otimes \mathbf{I}+\sigma_z\otimes \sigma_x+\sigma_z\otimes \sigma_y)\otimes \sigma_x$, where $\sigma_x$, $\sigma_y$ and $\sigma_z$ are the Pauli matrices. In this case, the orthogonal projections on the disjoint common eigenspaces of $H_0$, $L_1$ and $L_2$ are exactly the eight GHZ states for three-qubit systems. We determine $\frac12(|000\rangle+|111\rangle)(\langle000|+\langle 111|)$ as the target state. The values of the physical and experimental parameters are
chosen as $\omega = 0.3$, $\eta_1 = 0.5$, $\gamma_1=1.1$, $\eta_2=0.4$, $\gamma_2 = 1$ and $\hat{\omega} = 0.4$, $\hat{\eta}_1 = 0.54$, $\hat{\gamma}_1=1.05$, $\hat{\eta}_2=0.44$, $\hat{\gamma}_2 = 0.95$. We follow the approach in~\cite[Section 5]{cardona2020exponential} to construct the matrix $\Gamma$ based on $H_1$ appearing in the simplified filter~\eqref{Eq:ApproxSDE}.

\begin{figure}[!t]
\centerline{\includegraphics[width=10cm]{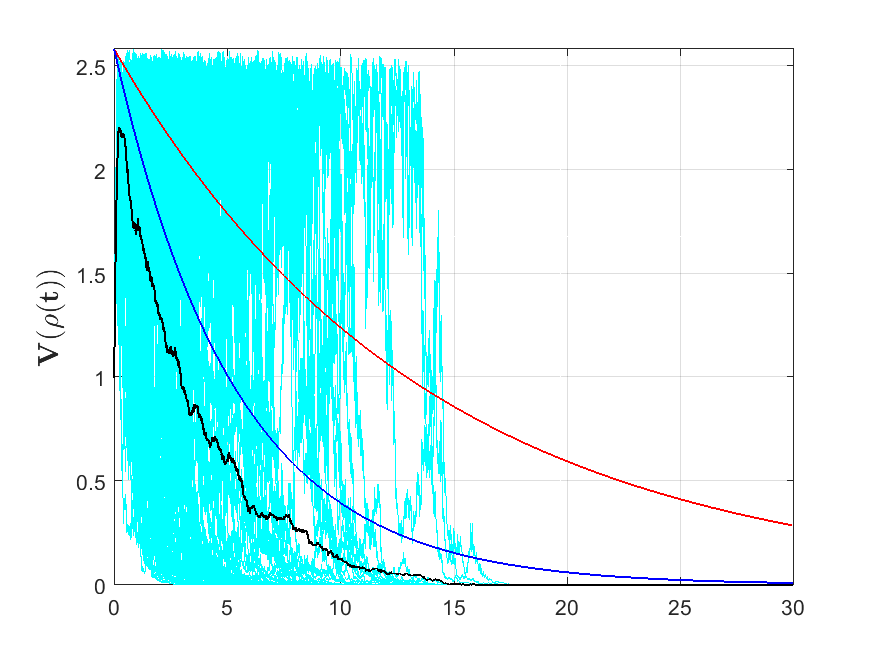}}
\caption{\small Exponential stabilization of the three-qubit system with feedback controller $u(\hat{\rho})$, $\rho(0)=\frac12(|011\rangle+|100\rangle)(\langle011|+\langle 100|)$ and $\hat{\rho}(0)=\mathbf{I}/8$. The blue curve represents the mean value of 100 realizations, the red curves represent the exponential reference with exponent $-0.0733$, the blue curve represents the exponential reference with exponent $-0.1876$.}
\label{Fig:FullFB}
\end{figure}

\begin{figure}[!t]
\centerline{\includegraphics[width=10cm]{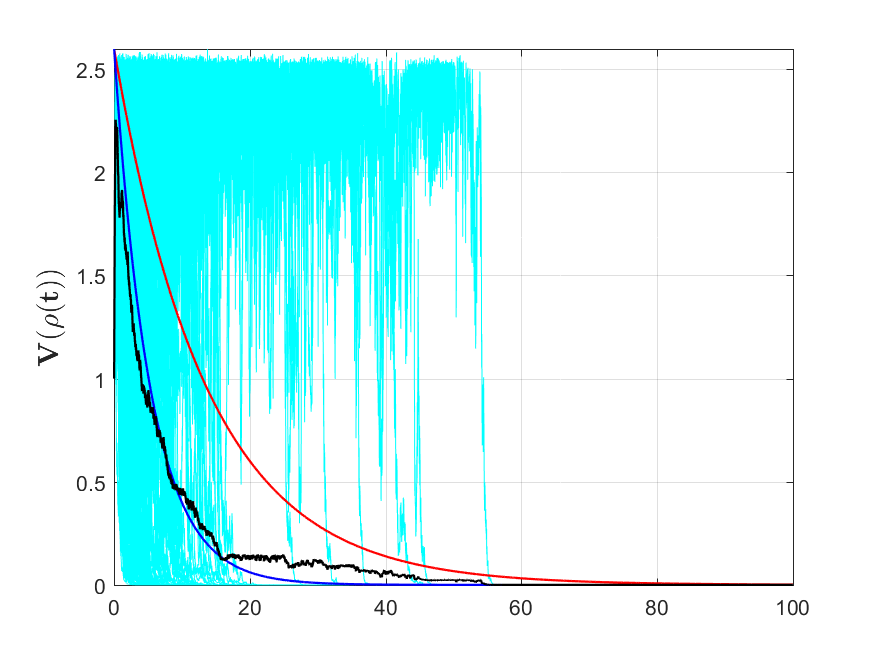}}
\caption{\small Exponential stabilization of the three-qubit system with feedback controller $u(\hat{q})$, $\rho(0)=\frac12(|011\rangle+|100\rangle)(\langle011|+\langle 100|)$ and $\hat{q}(0)=[1 1 1 1 1 1 1 1]^*/8$. The blue curve represents the mean value of 100 realizations, the red curves represent the exponential reference with exponent $-0.0733$, the blue curve represents the exponential reference with exponent $-0.1876$.}
\label{Fig:DiagFB}
\end{figure}

By straightforward calculation, the assumptions of Theorem~\ref{Thm:ExpStab} on control Hamiltonian and parameters are satisfied. We apply the feedback controller of the form~\eqref{Eq:u_General1} and~\eqref{Eq:u_General_q1} for the system which are illustrated in Fig.~\ref{Fig:FullFB} and Fig.~\ref{Fig:DiagFB} respectively, with $\alpha=5$, $\beta=2$, $\epsilon_1=0.1$ and $\epsilon_2=0.5$. 

Fig.~\ref{Fig:FullFB} and Fig.~\ref{Fig:DiagFB} show the behavior of the Lyapunov function considered in the proof of Theorem~\ref{Thm:ExpStab} along 100 sample trajectories, we observe that the simulated trajectories have an asymptotic behavior consistent with the estimated sample Lyapunov exponent $-C-K=-0.1876$ (blue curve). Nevertheless, Theorem~\ref{Thm:ExpStab} suggests that the rate of convergence of the expectation of the Lyapunov function should be less than or equal to $-C=-0.0733$ (red curve).

\section{Conclusion}
\label{sec:conc}
This paper shows the feedback stabilization of general open quantum systems undergoing QND measurements with different quantum channels, not necessarily Hermitian, toward a target subspace. This is of interest to stabilize entangled states which are of importance in quantum information processing. The robustness of the stabilizing feedback with respect to imperfections and an approximate filter constructed from the diagonal elements of the estimated filter, has been shown. Further investigations will be considered to stabilize generic open quantum systems with generic measurement operators, propose reduced filters and prove robustness of feedback to such reduced filters, and consider application in continuous-time quantum error corrections.

\appendix
\section{Stochastic tools}

\textbf{Infinitesimal generator and It\^o formula.}
Given a stochastic differential equation $dq(t)=f(q(t))dt+\sum^{m}_{k=1}g_k(q(t))dW_k(t)$, where $q$ takes values in $Q\subset \mathbb{R}^p$, then the infinitesimal generator is an operator $\mathcal{L}$ acting on the function $V: Q  \rightarrow \mathbb{R}$ which is twice continuously differentiable, in the following way
\begin{align}
&\mathscr{L}V(x):=\sum_{i=1}^p\frac{\partial V(x)}{\partial x_i}f_i(x,y)+\frac12 \sum^{m}_{k=1}\sum_{i,j=1}^p\frac{\partial^2 V(x)}{\partial x_i\partial x_j}(g_{k})_i(x)(g_{k})_j(x).
\label{Eq:InfinitesimalGenerator}    
\end{align}
It\^o formula describes the variation of the function $V$ along solutions of the stochastic differential equation and is given as follows
\begin{equation*}
\begin{split}
dV(q(t)) = &\mathscr{L}V(q(t))dt+\sum_{k=1}^m\sum_{i=1}^p\frac{\partial V(q(t))}{\partial q_i}(g_{k})_i(q(t))dW_k(t).
\end{split}
\end{equation*}

\textbf{Stratonovich equation and Support theorem.}
Any stochastic differential equation in It\^o form in $\mathbb R^K$
\begin{equation*}
dx(t)=\widehat X_0(x(t))dt+\sum^n_{k=1}\widehat X_k(x(t))dW_k(t), \quad x_0 = x,
\end{equation*}
can be written in the following Stratonovich form~\cite{rogers2000diffusions2}
\begin{equation*}
dx(t) = X_0(x(t))dt+\sum^n_{k=1}X_k(x(t)) \circ dW_k(t), \quad x(0) = x,
\end{equation*}
where 
$X_0(x)=\widehat X_0(x)-\frac{1}{2}\sum^K_{l=1}\sum^n_{k=1}\frac{\partial \widehat X_k}{\partial x_l}(x)(\widehat X_k)_l(x)$, $(\widehat X_k)_l$ denoting the component $l$ of the vector $\widehat X_k,$ and $X_k(x)=\widehat X_k(x)$ for $k\neq 0$.

\medskip
The following classical theorem relates the solutions of a stochastic differential equation with those of an associated deterministic one.
\begin{theorem}[Support theorem~\cite{stroock1972support}]
Let $X_0(t,x)$ be a bounded measurable function, uniformly Lipschitz continuous in $x$ and $X_k(t,x)$  be continuously differentiable in $t$ and twice continuously differentiable in $x$, with bounded derivatives, for $k\neq 0.$ Consider the Stratonovich equation
\begin{equation*}
dx(t) = X_0(t,x(t))dt+\sum^n_{k=1}X_k(t,x(t)) \circ dW_k(t),  x(0) = x,
\end{equation*}
and denote by $\mathbb{P}_x$ the probability law of the solution $x(t)$ starting at $x$. Consider in addition the associated deterministic control system
\begin{equation*}
\frac{d}{dt}x_{v}(t) = X_0(t,x_{v}(t))+\sum^n_{k=1}X_k(t,x_{v}(t))v_k(t), x_v(0) = x,
\end{equation*}
with $v_k \in \mathcal{V}$, where $\mathcal{V}$ is the set of all locally bounded measurable functions from $\mathbb{R}_+$ to $\mathbb{R}$. Define $\mathcal{W}_x$ as the set of all continuous paths from $\mathbb{R}_+$ to $\mathbb R^K$ starting at $x$, equipped with the topology of uniform convergence on compact sets, and $\mathcal{I}_x$ as the smallest closed subset of $\mathcal{W}_x$ such that $\mathbb{P}(x(\cdot) \in \mathcal{I}_x)=1$. Then, $\mathcal{I}_x = \overline{ \{ x_{v}(\cdot)\in\mathcal{W}_x|\, v \in \mathcal{V}^n\} } \subset \mathcal{W}_x.$
\label{Thm:Support}
\end{theorem}

\section{Invariance properties of trajectories}~\label{sec:pre}
Here, we state some fundamental results that are used in the proof of instability and recurrence. These results are analogous to the results in~\cite[Section~4]{liang2019exponential} for the coupled system~\eqref{Eq:SME_W}--\eqref{Eq:SME_filter_W}, and they concern invariance properties for the system, involving the boundary $\partial\mathcal{S}=\{\rho\in\mathcal{S}|\,\det({\rho})=0\}$ and the interior $\mathrm{int}(\mathcal{S})=\{\rho\in\mathcal{S}|\,{\rho}>0\}$.  Since their proofs are based on the same arguments, we omit them.
\begin{lemma}
Assume that \emph{\textbf{H0}} holds. Let $(\rho(t),\hat{\rho}(t))$ be a solution of~\eqref{Eq:SME_W}--\eqref{Eq:SME_filter_W}. If $\rho(0)\in\mathrm{int}(\mathcal{S})$, then 
$\mathbb{P}(\rho(t)\in\mathrm{int}(\mathcal{S}),\,\forall t\geq0)=1.$ 
Similarly, if $\hat{\rho}(0)\in\mathrm{int}(\mathcal{S})$, then $\mathbb{P}(\hat{\rho}(t)\in\mathrm{int}(\mathcal{S}),\,\forall t\geq0)=1.$ 
More in general, the ranks of ${\rho}(t)$ and $\hat{\rho}(t)$ are a.s. non-decreasing.
\label{Lemma:PosDef invariant}
\end{lemma}

\begin{lemma}
Assume that \emph{\textbf{H0}} holds.
If $\eta_k=1$ for all $k\in[M]$, then $\partial \mathcal{S}\times \mathcal{S}$ is a.s. invariant for the coupled system~\eqref{Eq:SME_W}--\eqref{Eq:SME_filter_W}. If $\hat{\eta}_k=1$ for all $k\in[M]$, then $\mathcal{S}\times\partial\mathcal{S}$ is a.s. invariant for the coupled system~\eqref{Eq:SME_W}--\eqref{Eq:SME_filter_W}.
\label{Lemma:Boundary invariant}
\end{lemma}

\begin{lemma}
Assume that \emph{\textbf{H0}} is satisfied. Let $(\rho(t),\hat{\rho}(t))$ be a solution of the coupled system~\eqref{Eq:SME_W}--\eqref{Eq:SME_filter_W}. If
$\hat{\rho}(0) \notin \mathcal{I}(\mathcal{H}_{\bar{n}})$, then $\mathbb{P}( \hat{\rho}(t) \notin \mathcal{I}(\mathcal{H}_{\bar{n}}), \forall\, t\geq 0 )=1.$
Moreover, if $\hat{\rho}(0)\in \mathcal{I}(\mathcal{H}_{\bar{n}})$ and $\rho(0)\notin \mathcal{I}(\mathcal{H}_{n})$, then $\mathbb{P}( \rho(t) \notin \mathcal{I}(\mathcal{H}_{n}), \forall\, t\geq 0 )=1$ for any $n\in[M]$. 
\label{Lemma:NeverReach}
\end{lemma}

\section{Exiting from boundary}

Next, we introduce the following assumption,
\smallskip
\begin{itemize}
\item[\textbf{H4}:] 
$\{\rho\in\mathcal{S}|\,u(\rho)=0\}$ does not contain the complete integral curves of $\tilde{\mathcal{L}}^0_{\hat{\omega},\hat{\gamma},\hat{\eta}}(\hat{\rho}_v)$ and $\mathcal{G}^k_{\hat{\eta}_k,\hat{\gamma}_k}(\hat{\rho}_v)$ for $k\in[M]$.
\end{itemize}
\smallskip
\begin{proposition}
Suppose that \emph{\textbf{H0}}, \emph{\textbf{H3}} and \emph{\textbf{H4}} hold true, and there exists at least one $k\in[M]$ such that $\eta_k,\hat{\eta}_k\in(0,1)$. 
Then, for~\eqref{Eq:ODE}--\eqref{Eq:ODE_F} and any $v(t)\in\mathcal{V}^m$, there exits a finite constant $T_v\geq 0$ such that $\rho_v(t)>0$ and $\hat{\rho}_v(t)>0$ for all $t>T_v$.
\label{Prop:ExitBoundary}
\end{proposition}

\proof
Due to \hyperref[Lemma:PosDef invariant]{Lemma~\ref*{Lemma:PosDef invariant}} and the support theorem (\hyperref[Thm:Support]{Theorem~\ref*{Thm:Support}}), if $\hat{\rho}(0)>0$, then $\hat{\rho}(t)>0$ for all $t\geq 0$ and trajectories of~\eqref{Eq:ODE_F} stay in $\mathrm{int}(\mathcal{S})$ (the same arguments can be applied for $\rho$). Now, we consider the case $(\rho(0),\hat{\rho}(0))\in(\partial\mathcal{S}\times \partial\mathcal{S})\setminus (\mathcal{I}(\mathcal{H}_{\bar{n}})\times\mathcal{I}(\mathcal{H}_{\bar{n}}))$. Here, we only focus on the case of $\hat{\rho}_v(t)$. The case of $\rho_v(t)$ can be proved in the same manner. 
We follow the $Z_1$-$Z_2$ arguments in~\cite[Proposition 4.5]{liang2019exponential}. Define 
\begin{equation*}
\begin{split}
&\widehat{Z}_1(t):=\mathrm{span}\{\phi_k\in\mathscr{B}|\,\phi_k^*\hat{\rho}_v(t)\phi_k=0\},\\
&\widehat{Z}_2(t):=\text{eigenspace related to the eigenvalue 0 of } \hat{\rho}_v(t).
\end{split}
\end{equation*}
By definition, $\widehat{Z}_1(t)\subseteq\widehat{Z}_2(t)$ for all $t\geq 0$, and $\widehat{Z}_1(t)$ is the largest subspace of $\widehat{Z}_2(t)$ which is invariant by $L_1,\dots,L_m$, since the measurement operators are diagonal in $\mathscr{B}$.

Denote by $\hat{\lambda}_k(t)$ and $\hat{\psi}_k(t)$ for $k\in\{1,\dots,N\}$ the eigenvalues and eigenvectors of $\hat{\rho}_v(t)$. Due to Kato theorem (\cite[Theorem 2.6.8]{kato1976perturbation}), $\hat{\lambda}_k(t)\in\mathcal{C}^1$ since $\hat{\rho}_v(t)\in\mathcal{C}^1$. Inspired by Feynman-Hellmann theorem, for any $\hat{\psi}_j(t)\in \widehat{Z}_2(t)$ for $t\in[0,\varepsilon]$ with $\varepsilon>0$ sufficiently small, the corresponding eigenvalue satisfies
\begin{equation*}
\dot{\hat{\lambda}}_j(t)=\sum^m_{j=1}\hat{\gamma}_k(1-\hat{\eta}_k)\hat{\psi}^*_j(t)L_k\hat{\rho}_v(t)L_k^*\hat{\psi}_j(t).
\end{equation*}
If $\hat{\psi}_j(t)\in\widehat{Z}_2(t)\setminus \widehat{Z}_1(t)$, then $L^*_k\hat{\psi}_j(t)\notin \widehat{Z}_2(t)$ for all $k\in[M]$, since otherwise $\widehat{Z}_1(t)$ would not be the largest subspace invariant by $L_1,\dots,L_m$ contained in $\widehat{Z}_2(t)$. By the similar arguments as in~\cite[Proposition 4.5]{liang2019exponential}, we have $\widehat{Z}_1(s)\subseteq\widehat{Z}_2(s)\subseteq\widehat{Z}_1(t)$ for all $s-t>0$ sufficiently small. Moreover, due to the assumptions on feedback controller, for each $v(t)\in\mathcal{V}^m$, there exists a finite constant $T_v\geq 0$ such that $u(\hat{\rho}_v(t))\neq 0$ for all $t\in[T_v,T_v+\varepsilon]$ with $\varepsilon>0$ sufficiently small. By employing the similar arguments as in~\cite[Lemma 9]{liang2022GHZ}, for all $\xi\in\widehat{Z}_1(T_v)$, $\xi\notin\widehat{Z}_1(t)$ for $t-T_v>0$ arbitrarily small. Then the proposition can be concluded by following the same arguments as in~\cite[Proposition 4.5]{liang2019exponential}.\hfill$\square$

The following corollary of \hyperref[Prop:ExitBoundary]{Proposition~\ref*{Prop:ExitBoundary}} is a direct application of~\cite[Theorem 2.3.2]{shmatkov2006rate} and~\cite{gyongy2004wong}, in which the authors showed the convergence rate of Wong-Zakai approximation for a type of stochastic differential equations and then inferred the corresponding almost sure convergence (see also~\cite{brzezniak1995almost} for another treatment). 
\begin{corollary}
Suppose that the hypotheses in \hyperref[Prop:ExitBoundary]{Proposition~\ref*{Prop:ExitBoundary}} are satisfied. Then, for the coupled system~\eqref{Eq:SME_W}--\eqref{Eq:SME_filter_W}, for almost all $\omega\in\Omega$, there exists $T(\omega)<\infty$ such that $\rho(t)>0$ and $\hat{\rho}(t)>0$ for all $t>T(\omega)$.
\label{Cor:ExitBoundary}
\end{corollary}

\section{Acknowledgements}
This work is partially supported by the Agence Nationale de la Recherche projects Q-COAST ANR- 19-CE48-0003 and IGNITION ANR-21-CE47-0015, and by the Department of Information Engineering of the University of Padova. W.L. thanks Dr. Chenmin Sun for stimulating discussion on the instability and Wong-Zakai approximation. The authors also thank Dr. Paolo Mason and Mario Neufcourt for the helpful discussion.

\bibliographystyle{plain}        
\bibliography{Ref_ReducedFeedback}           

\end{document}